\documentclass[reprint,pra,superscriptaddress,twocolumn,showpacs]{revtex4-1}
\usepackage{graphicx}
\usepackage{amsmath}
\usepackage{amsfonts}
\usepackage{bm}
\usepackage{color}
\usepackage{cancel}

\usepackage[plainpages=false,pdfpagelabels,colorlinks=true,linkcolor=red,urlcolor=blue,citecolor=blue,pdftitle={Title},pdfauthor={},pdfdisplaydoctitle=true,pdfduplex=DuplexFlipLongEdge]{hyperref}

\definecolor{darkred}{rgb}{0.10,0.10,0.80}

\newcommand{\ket}[1]{|#1\rangle}

\begin{document}
\title{
Thermalization after photoexcitation from the perspective of optical spectroscopy 
}

\author{Jan \surname{Kogoj}}
\affiliation{J. Stefan Institute, 1000 Ljubljana, Slovenia}

\author{Lev \surname{Vidmar}}
\affiliation{Department of Physics and Arnold Sommerfeld Center for Theoretical Physics, Ludwig-Maximilians-Universit\"at M\"unchen, D-80333 M\"unchen, Germany}
\affiliation{J. Stefan Institute, 1000 Ljubljana, Slovenia}

\author{Marcin \surname{Mierzejewski}}
\affiliation{Institute of Physics, University of Silesia, 40-007 Katowice, Poland}

\author{Stuart A. \surname{Trugman}}
\affiliation{Center for Integrated Nanotechology, Los Alamos National Laboratory, Los Alamos, New Mexico, USA}

\author{Janez \surname{Bon\v{c}a}}
\affiliation{J. Stefan Institute, 1000 Ljubljana, Slovenia}
\affiliation{Faculty of Mathematics and Physics, University of Ljubljana, 1000
Ljubljana, Slovenia}

\begin{abstract}
We analyze the thermalization of a photoexcited charge carrier coupled to a single branch of quantum phonons within the Holstein model.
To this end, we calculate the far-from-equilibrium time evolution of a pure many-body state and compare it with predictions of the thermal Gibbs ensemble.
We show that at strong enough carrier excitation, the nonequilibrium system evolves towards a thermal steady state.
Our analysis is based on two classes of observables.
First, the occupations of fermionic momenta, which are the eigenvalues of the one-particle density matrix, match in the steady state the values in the corresponding Gibbs ensemble.
This indicates thermalization of static fermionic correlations on the entire lattice.
Second, the dynamic current-current correlations, including the time-resolved optical conductivity, also take the form of their thermal counterparts.
Remarkably, both static and dynamic fermionic correlations thermalize with identical temperatures.
Our results suggest that the subsequent relaxation processes, observed in time-resolved ultrafast spectroscopy, may be efficiently described by applying quasithermal approaches, e.g., multi-temperature models. 
\end{abstract}
\maketitle

\section{Introduction}

Equilibration and thermalization of closed quantum many-body systems~\cite{polkovnikov11,eisert15,dalessio_kafri_15} are central topics in a broad and very active field of quantum physics far from equilibrium.
Generic systems are expected to thermalize on a level of eigenstates~\cite{rigol08,dalessio_kafri_15} and the reduced density matrices of their subsystems approach the Gibbs form~\cite{Linden2009}.
In solids, charge carriers are always coupled to other degrees of freedom that represent their environment.
This assures that, at sufficiently long times after perturbation, the system thermalizes.

Time-resolved optical experiments may, however, photoexcite charge carriers on extremely short time scales~\cite{giannetti16,dalconte15,okamoto2010,gadermaier10,dalconte12,gadermaier14,novelli14}.
A state-of-the-art example is the recent work by Dal Conte {\it et al}~\cite{dalconte15} that studied the time evolution of a cuprate superconductor, photoexcited and probed within 10-20~fs.
This time scale, which is only a fraction of a typical phonon oscillation period, implies that the photoexcitation can be, to a good approximation, considered as a quantum quench, generating nonthermal states of photoexcited carriers on short times.

The dynamics in a short time window is usually governed by excitations that couple most strongly to  carriers
(we denote this time regime as primary relaxation).
On the other hand, the full equilibration process requires redistribution of the excess energy over the whole environment, which is significantly slower than the primary relaxation time.
A natural question thus arises, at which time scale is one justified in describing the dynamics by using theories that rely on the assumption of (quasi)equilibrium distributions (e.g., the so-called multi-temperature models~\cite{kaganov1957,allen87}), and on which time scale are genuinely nonequilibrium concepts required?

The part of the environment responsible for the primary relaxation may depend on a particular material.
The dominant relaxation channels are usually phonons~\cite{golez_prl2012,defilippis12,matsueda12,kennes10,kemper13,sentef13,werner2013,hohenadler13,baranov14,kemper14,werner15,aoki2015,lev2015,sayyad15,mishchenko15,werner16,rizzi15}, bosons of electronic origin~\cite{golez2014,bonca2012,iyoda14,eckstein14b,dalconte15}, a combination of the latter two~\cite{lev2011,kogoj2014} or a direct carrier-carrier interaction~\cite{eckstein2013,moritz13,rincon14,werner14}.
In this work, we are interested in systems where the primary relaxation channel of photoexcited electrons are bosons of local origin.
The relevance of such a relaxation mechanism has been highlighted, e.g., in pump-probe experiments on optimally doped cuprates~\cite{dalconte12,dalconte15}.

The goal of this work is to study efficiency of thermalization in the primary relaxation regime, which is modeled by a closed quantum system of a single electron (described by $\hat H_{\rm kin}$)
coupled to bosons.
For simplicity, we model the bosons as optical phonons (described by $\hat H_{\rm ph}$) and study the Holstein model with the general form
\begin{equation} \label{eqHtot}
\hat H_{\rm sys} = \hat H_{\rm kin} + \hat H_{\rm ph} + \hat H_{\rm e-ph},
\end{equation}
where the last term is the electron-phonon interaction
[we define the terms of Eq.~(\ref{eqHtot}) below].

We contrast the outcome of unitary quantum time evolution with predictions of statistical mechanics.
If the electron excitation is strong enough, the system approaches a steady state.
Observables in the steady state are, at a given total energy, independent of initial conditions.
We show that the eigenvalues of the electronic one-particle density matrix (i.e., the occupations of fermionic momenta) approach the ones in the Gibbs ensemble.
This result is consistent with thermalization of static electronic observables.
We test thermalization separately
using the  nonequilibrium optical conductivity (and more generally, dynamical current-current correlations), which in the steady state as well approach a  thermal form. 
Our results indicate that, for photoexcited systems at small photodoping, it is very likely that the subsequent (secondary) relaxation processes can be efficiently described by applying quasithermal approaches.

The paper is organized as follows.
In Sec.~\ref{sec2} we introduce the model, numerical methods, and the quench protocols to study relaxation and thermalization.
Thermalization of electronic one-particle observables is studied in Sec.~\ref{sec3}.
In Sec.~\ref{sec4} we study dynamical correlation functions and their approach to a thermal form.
We conclude in Sec.~\ref{sec5}.
Details of the numerical methods and definitions of dynamical correlation functions are given in Appendices~\ref{sec:numerics}-\ref{ap:FTLM} and~\ref{sp:com}-\ref{sp:sigom}, respectively.

\section{Model and quench protocols} \label{sec2}

We consider an electron on a periodic one-dimensional lattice with $L$ sites
\begin{equation} \label{Hele}
\hat H_{\rm kin} = -t_{0}\sum_{j} \left({e^{i \phi(t)}} \hat c_{j}^{\dagger}  \hat c_{j+1} + \mbox{h.c.} \right) = \sum_k \epsilon_{k(t)} \hat c_k^\dagger \hat c_k,
\end{equation}
where $t_{0}$ is the nearest-neighbor hopping amplitude and  $\hat c_{j}$ is a fermion annihilation operator on site $j$.
We use the Peierls substitution to couple the electron to an external dimensionless electric field $F$ (related to the electric field $\cal F$ as $F = e_0 {\cal F} a/t_0$, where $e_0$ is the unit charge and $a$ is the lattice constant~\cite{mierzejewski2011})  by introducing the time-dependent phase $\phi(t)=-(t/\tau) F $.
We measure time in units of $\tau = \hbar/t_0$,  and set $\hbar=1$ throughout the paper.
The momentum-space representation is evoked by a discrete Fourier transformation
$\hat c_k = \sum_j e^{-ijak} \hat c_j / \sqrt{L}$
leading to
$\epsilon_{k(t)} = -2t_0 \cos{(k a-\phi(t))}$.
The lattice constant $a$ is set to unity hereafter.
Quantum phonons are described by a single frequency $\omega_0$,
\begin{equation} \label{Hph}
\hat H_{\rm ph} = \omega_0 \sum_{j} \hat b_{j}^\dagger  \hat b_{j},
\end{equation}
where $\hat b_{j}$ is the phonon annihilation operator.
The electron is coupled to phonons via a Holstein-type interaction of strength $g$,
\begin{equation} \label{Heph}
\hat H_{\rm e-ph} = -g \sum_{j} \hat c_{j}^{\dagger}\hat c_{j} (\hat b_{j}^\dagger + \hat b_{j}).
\end{equation}
If not specified otherwise, we set $L=16$.
The dimensionless electron-phonon coupling is given by $\lambda=g^2/(2t_0\omega_0)$.

We construct two qualitatively distinct initial states, which are not eigenstates of $\hat H_{\rm sys}$.
In the first case, the initial state at $t=0$ consists of an electron at momentum $k$=$\pi$ and no phonons.
By suddenly switching on a finite value of electron-phonon interaction, we time evolve the system under $\hat H_{\rm sys}$.
Here, $\phi(t)$=0 for all times.
Such a state initially possesses the maximal electron kinetic energy
$E_\mathrm{kin}^{\rm}(t$=$0)=\langle \hat H_{\mathrm{kin}}(t$=$0)\rangle =2 t_0$,
which equals the total energy $E_{\rm sys} = \langle \hat H_{\rm sys} \rangle$.
We refer to this setup as the {\it interaction} quench, and the relaxation dynamics emerging from this initial state has been recently studied in~\cite{lev2015}.

The second initial state is prepared by switching on a uniform dc electric field $F$ in a time window $[-t_{\rm i},0]$.
We refer to this setup as the {\it field} quench.
At $t = -t_{\rm i}$, we set $\phi(t$=$-t_{\rm i})=0$, switch on $F$ and propagate  the ground state of $\hat H_{\rm sys}$ with $\phi(t)=-[(t+t_{\rm i})/\tau] F$ at chosen $\lambda$, $\omega_0$ and the total momentum $K$=$0$~\cite{lev2011_1,golez2012}.
We choose $t_{\rm i}$ such that at $t$=$0$ when $F$ is switched off, the total energy reaches a desired targeted value $E_\mathrm{sys}$.
The latter state is our initial state and we study its equilibration at $t>0$ in the absence of electric field, i.e., at the constant value $\phi(t>0)=-(t_{\rm i}/\tau) F$.
The inset of Fig.~\ref{fig1}(b) displays an example where the system is driven in the time interval $[-t_{\rm i},0]$ by a dc electric field $F=1.0$, until the target value $E_{\rm sys}=2t_0$ is reached.

In all cases under consideration, the system energy is conserved for $t>0$ ($\hat H_{\rm sys}$ is time independent) and the dynamics is described by the pure state $| \psi (t) \rangle = e^{-i \hat H_{\rm sys} t} |\psi_0 \rangle$, where $|\psi_0 \rangle$ is the initial state.
Results from the time propagation are compared with the predictions of the thermal Gibbs ensemble, described by the density matrix
\begin{equation} \label{def_Gibbs}
\hat \rho_{\rm sys} = Z^{-1}\exp[- \hat H_{\rm sys}/(k_{\rm B}T)],
\end{equation}
with the partition function $Z = {\rm Tr}\{  \exp[- \hat H_{\rm sys}/(k_{\rm B}T)]\}$.
We set the Boltzmann constant $k_\mathrm{B}=1$ hereafter and measure $T$ in units of $t_0$.

We apply the Lanczos-based diagonalization~\cite{park1986} in a limited functional space (LFS) for the ground-state calculation, the time evolution as well as to calculate equilibrium properties at finite temperature.
The generation of the LFS is described in Appendix~\ref{sec:numerics}.
It efficiently selects states with different phonon configurations around the electron~\cite{bonca99}, and is well-designed to describe systems with large phonon fluctuations that are not accessible by exact diagonalization.
The ground-state properties of the Holstein polaron have been calculated within the LFS with high numerical precision~\cite{bonca99,ku2002}.
The method has been recently extended to treat nonequilibrium problems~\cite{lev2011_1} 
and was shown to provide numerically precise quantum evolution of a closed many-body system~\cite{lev2015}.
In this work, we also use the LFS to calculate equilibrium properties in the Gibbs ensemble~(\ref{def_Gibbs}) using the Finite-Temperature Lanczos method (FTLM)~\cite{jaklic00}.
In Appendix~\ref{ap:FTLM}, we present details of the latter method as well as study finite-size effects in Fig.~\ref{figS4}.

\section{Electronic subsystem} \label{sec3}

\subsection{Kinetic energy}

\begin{figure} 
\includegraphics[width=0.43\textwidth]{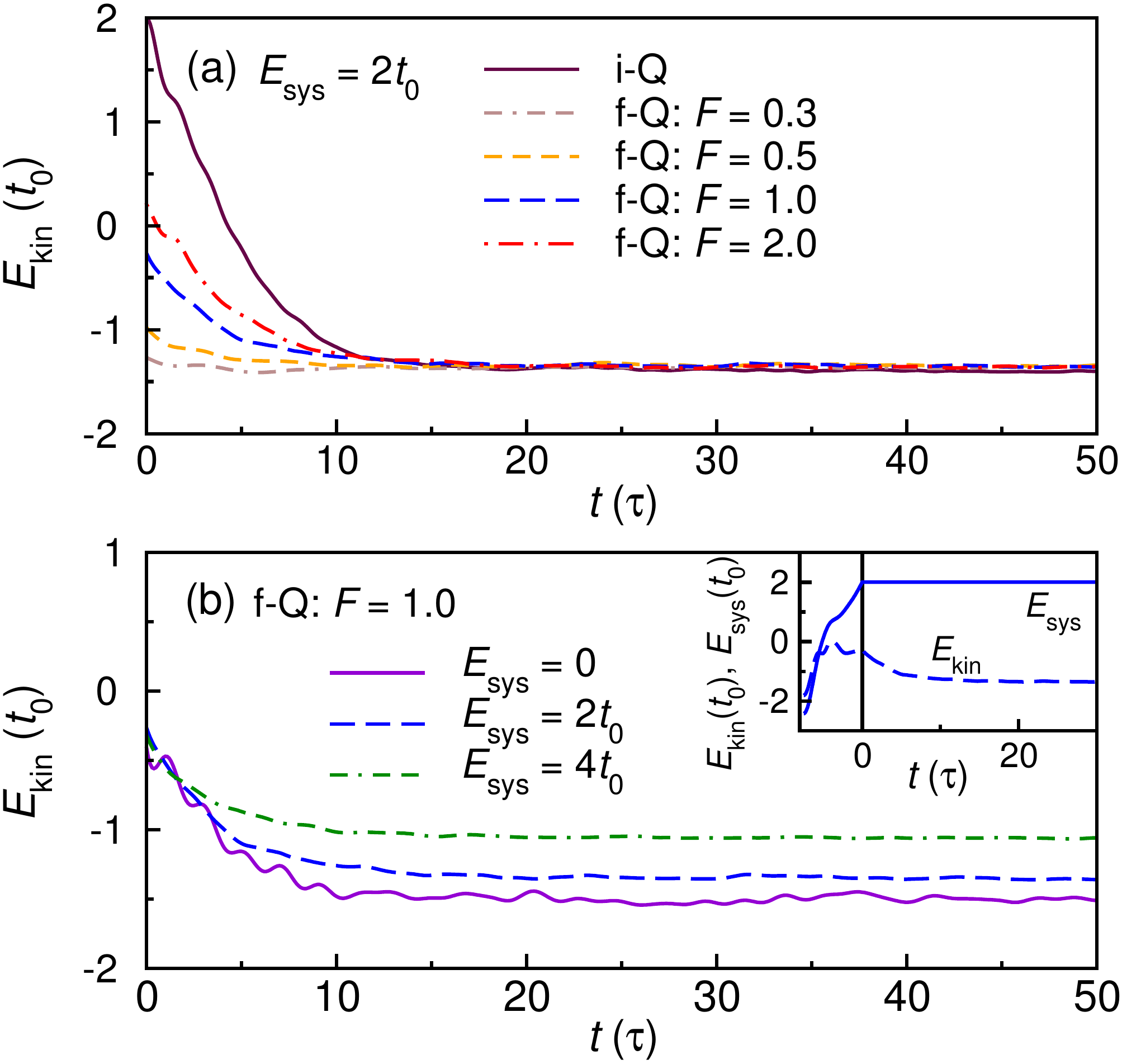}
\caption{
Time evolution of the electron kinetic energy at $\lambda=0.5$ and $\omega_0/t_0=0.75$.
(a) $E_\mathrm{kin}$ for different initial states and the same total energy $E_{\rm sys}=2t_0$.
The quench protocols are the interaction quench (i-Q) and the field quench (f-Q) for different values of the electric field $F$.
(b) $E_\mathrm{kin}$ for the field quench at $F=1.0$, corresponding to different total energies $E_{\rm sys}$.
Inset: $E_\mathrm{kin}$ and $E_{\rm sys}$ versus $t$ for the field quench at $F=1.0$.
The data are shown during the driving ($t<0$) and during equilibration ($t>0$), such that at $t=0$ we reach the target energy $E_{\rm sys}=2t_0$.
}
\label{fig1}
\end{figure}

We first study equilibration of the electron kinetic energy $E_{\rm kin}$ for different initial states and electron-phonon coupling strengths.
Figure~\ref{fig1}(a) shows the time evolution at $\lambda=0.5$ and $\omega_0/t_0=0.75$ after distinct quench protocols that  generate the same energy $E_{\rm sys}=2t_0$.
We observe   in the steady state a nearly indistinguishable value of $E_{\rm kin}\simeq -1.35t_0$.
Other observables such as $\langle \hat H_{\rm ph}\rangle$ and $\langle \hat H_{\rm e-ph}\rangle$ show qualitatively the same behavior (not shown here).
In the steady state and for the parameter regime under investigation, tiny oscillations around the average vanish in the limit $L \to \infty$~\cite{lev2015}.

On the other hand, quench protocols that  generate distinct total energy lead to distinct steady-state  values.
Figure~\ref{fig1}(b) shows results for the field quench with $F=1.0$, but distinct target energies $E_{\rm sys}=0,2t_0,4t_0$.
Interestingly, even though in all cases $E_{\rm kin}$ at $t=0$ is almost identical, the curves clearly deviate at later times.

\begin{figure} 
\includegraphics[width=0.43\textwidth]{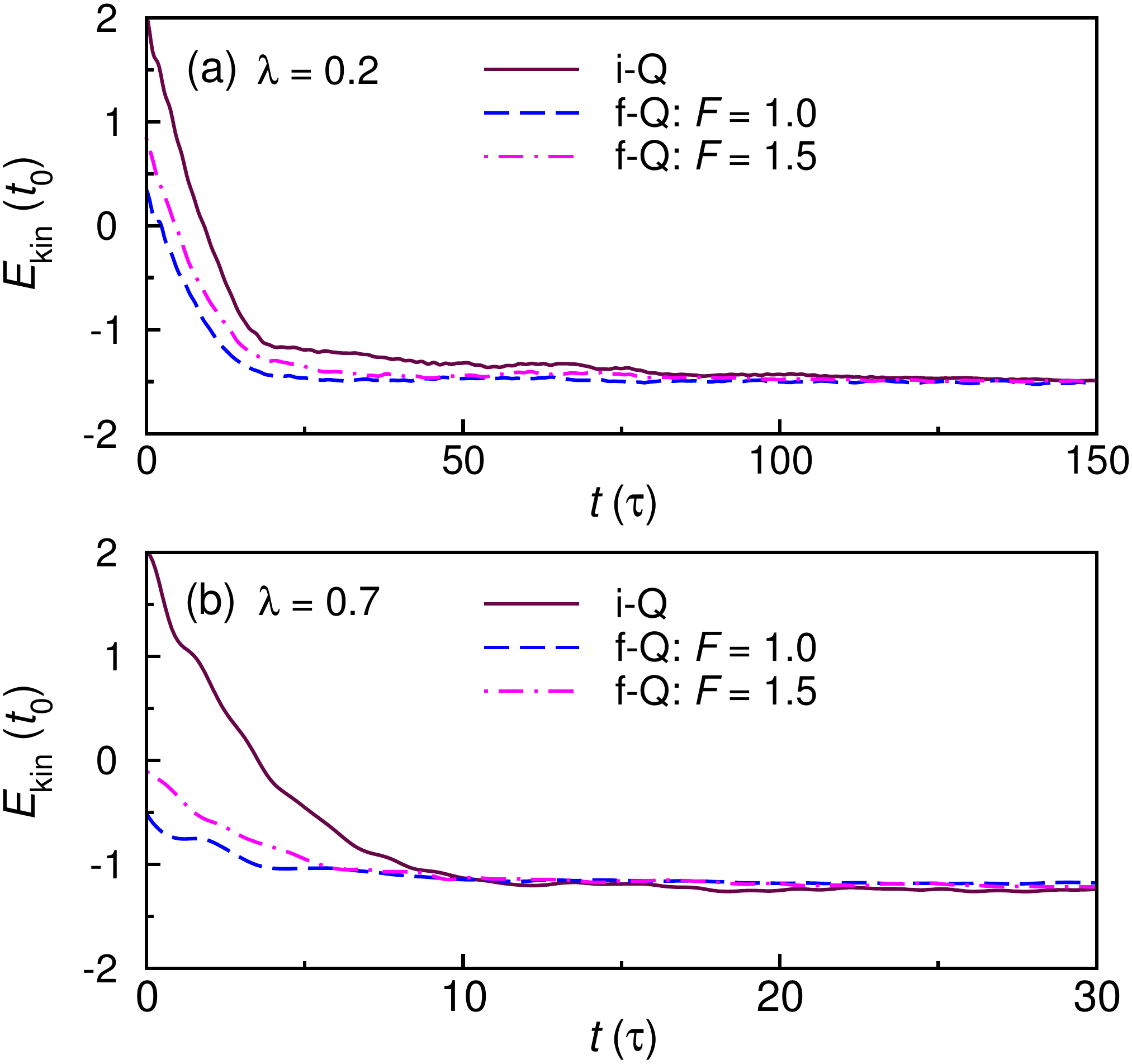}
\caption{
Time evolution of the electron kinetic energy  $E_\mathrm{kin}$ at $\omega_0/t_0=0.75$ and two values of the electron-phonon coupling strength
$\lambda=0.2$ (a) and $\lambda=0.7$ (b).
The quench protocols are the interaction quench (i-Q) and the field quench (f-Q) for two values of the electric field $F=1.0$ and $1.5$.
The total energy is $E_{\rm sys}=2t_0$ in both panels.
}
\label{fig1b}
\end{figure}

If the observed collapse of the data in Fig.~\ref{fig1}(a) for different initial states and the same total energy is generic, it should also be observed for other parameters of the model.
In Figs.~\ref{fig1b}(a) and~\ref{fig1b}(b) we show results for the electron kinetic energy at two other values of the electron-phonon coupling $\lambda=0.2$ and $0.7$, respectively.
When the steady state is reached, the data approach the same value, in a similar manner as in Fig.~\ref{fig1}(a) for $\lambda=0.5$.
This indicates that at both weak and moderate couplings, observables in the steady states may be characterized only by the total energy (and, as discussed in the following, described by the Gibbs ensemble $\hat \rho_{\rm sys}$).
It is also well known that the Holstein model exhibits the strong-coupling regime for $\lambda \gtrsim 1$, where a narrow polaronic band induces gaps in the many-body spectrum.
This does not necessarily exclude a possible thermal character of the steady states.
Nevertheless, large phonon fluctuations in the latter regime make the numerical comparison between the nonequilibrium and thermal equilibrium properties less accurate and will not be discussed in detail here.
Another interesting feature of the model is the influence of the phonon energy $\omega_0$.
We are going to discuss differences between the cases $\omega_0 < t_0$ and $\omega_0 > t_0$ in Sec.~\ref{subsec_opdm}.

The dynamics of $E_{\rm kin}$ in the weak-coupling regime displays two distinct time regimes, see Fig.~\ref{fig1b}(a).
After an initial fast decay, a slowing down in relaxation is observed when the electron kinetic energy becomes less than $\omega_0$ above the long-time steady-state value.
This effect has already been observed in single-electron~\cite{golez_prl2012} and many-electron~\cite{sentef13} systems coupled to optical phonons.
In contrast, such distinction of time regimes is less obvious for intermediate electron-phonon couplings shown in Figs.~\ref{fig1}(a) and~\ref{fig1b}(b), where the characteristic times
to reach the steady state become shorter~\cite{lev2015,sayyad15}.
This regime is going to be the main focus in the rest of the study.

\subsection{One-particle density matrix} \label{subsec_opdm}

To gain additional insight into the steady-state properties beyond the information gained from calculating a few selected observables~\cite{our2013,sedlmayr13,fagotti13}, we calculate the reduced density matrix of a small subsystem.
We focus on $\lambda=0.5$ furtheron.
In the Holstein model, it is a straightforward choice to take the subsystem consisting of electronic degrees of freedom only, while phonon degrees of freedom act as an environment.
The reduced density matrix is hence defined as $\hat \rho_{\rm ele} = {\rm Tr}_{\rm ph} \{\hat \rho_{\rm tot} \}$, where $\hat \rho_{\rm tot}$ denotes the density matrix of the total system.
(Note that $\hat \rho_{\rm tot}$ may be obtained from the pure state, $\hat \rho_{\rm tot} = | \psi(t) \rangle \langle \psi(t) |$, or from the Gibbs ensemble, $\hat \rho_{\rm tot} = \hat \rho_{\rm sys}$.)
This selection of $\hat \rho_{\rm ele}$ is different from the one where the subsystem is physically separated from the environment.
In the case of a single electron studied here, $\hat \rho_{\rm ele}$ is equivalent to the one-particle density matrix with the matrix elements $\hat c_j^\dagger \hat c_l$.
Due to translational invariance, $\hat \rho_{\rm ele}$ is diagonal in the momentum representation and its eigenvalues are occupations of the momentum states $\langle k | \hat \rho_{\rm ele} | q \rangle = n_k \delta_{k,q}$.
In the nonequilibrium calculation, we therefore need to calculate the average value of $n_k(t)$ in the steady  state $\bar n_k = \langle n_k(t) \rangle_t$,
where
\begin{equation}
n_k(t) = 1/L \sum_{j,l} e^{-i(l-j)k}  \langle \psi(t) | \hat c_j^\dagger \hat c_l | \psi(t) \rangle
\end{equation}
is the time-dependent momentum distribution function and $\langle ... \rangle_t$ denotes the time averaging.

\begin{figure}
\includegraphics[width=0.47\textwidth]{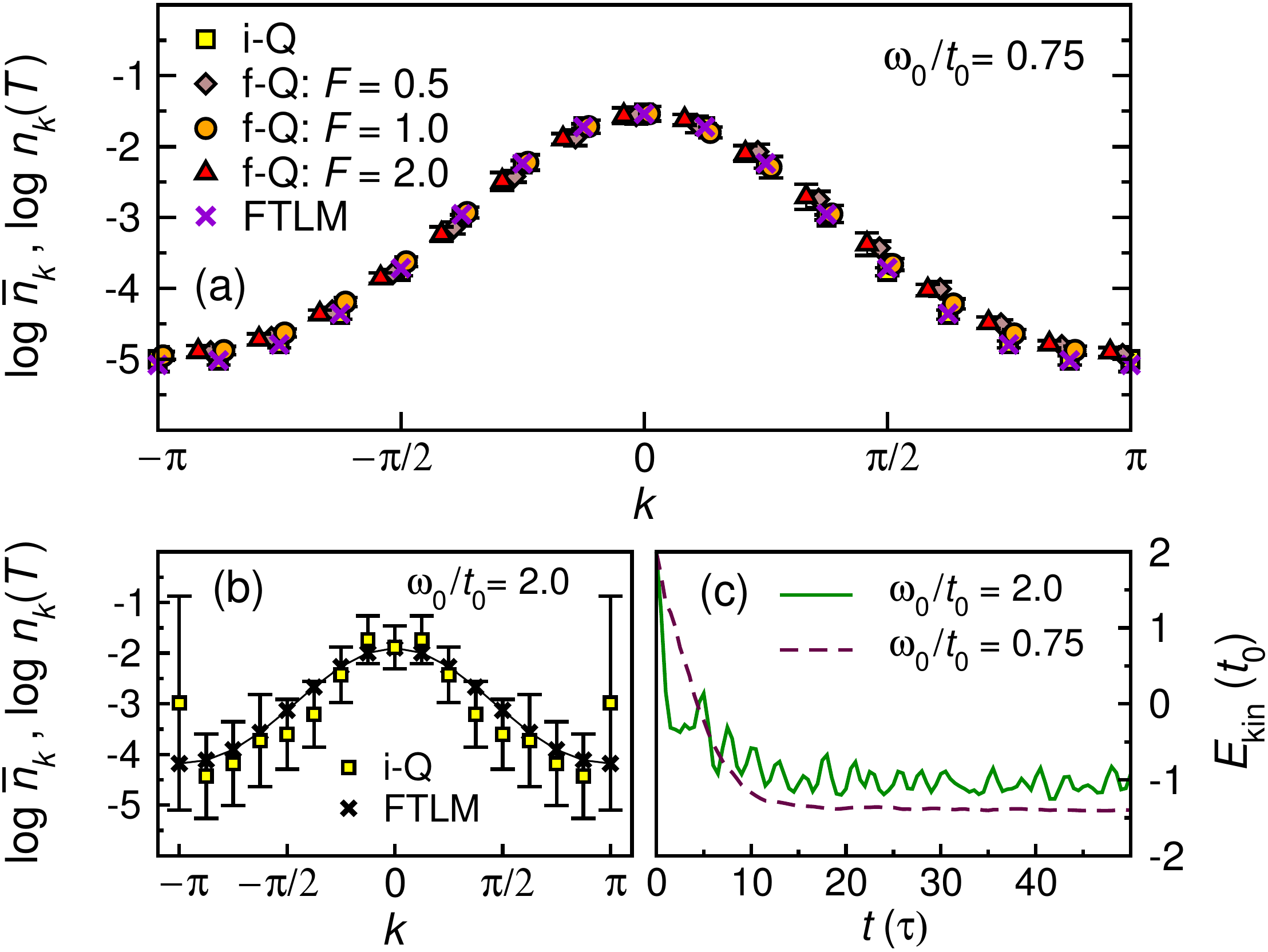}
\caption{
(a) and (b):
Eigenvalues of the one-particle density matrix $\langle k | \hat \rho_{\rm ele} | k \rangle = n_k$ at the total energy $E_{\rm sys}=2t_0$ and $\lambda=0.5$.
For the time-evolved pure state, we plot $\log \bar n_k$ for different initial states.
The data are averaged in the time interval $t \in [50\tau,100\tau] $.
Error bars equal three times the temporal standard deviation.
The thermal prediction $\log  n_k(T)$ is obtained from the Gibbs ensemble~(\ref{def_Gibbs}).
(a)
Results for $\omega_0/t_0=0.75$.
We compare $\log \bar n_k$ after the interaction quench (i-Q) and the field quench (f-Q), with $\log  n_k(T)$ at temperature $T=0.68$. 
For the field quench, we shift $k \to k-\phi_0$, where $\phi_0 = -(t_{\rm i}/\tau) F$ and $t_{\rm i}$ defines the time interval when $F$ is switched on, as explained in Sec.~\ref{sec2}.
(b)
Results for $\omega_0/t_0=2.0$.
We compare  $\log \bar n_k$  after the interaction quench with $\log  n_k(T)$ at $T = 1.36$.
(c) $E_{\rm kin}$ versus $t$ after the interaction quench for $\omega_0/t_0=2.0$ and $0.75$.
}
\label{fig2}
\end{figure}

A necessary criterion for thermalization requires a single set of $\bar n_k$ for all initial states with equal energy~\cite{Linden2009}.
In Fig.~\ref{fig2}(a) we plot $\log \bar n_k$ for different initial states at the energy $E_{\rm sys}=2t_0$.
We observe a collapse of the data in the entire Brillouin zone.  
 
We also present a comparison of $\log \bar n_k$ with the equilibrium results  $\log  n_k(T)$ in the Gibbs ensemble~(\ref{def_Gibbs}) obtained using the FTLM.
We have chosen the time-averaged electron kinetic energy in the steady state $\langle E_{\rm kin} \rangle_t$ to determine the temperature $T$ in the Gibbs ensemble through the equation
$\langle E_{\rm kin} \rangle_t = {\rm Tr} \{ \hat \rho_{\rm sys} \hat H_{\rm kin} \}$.
(The reason for this choice is that within the FTLM, finite-size effects due to the limited number of phonon configurations affect $E_{\rm kin}$ less than $E_{\rm sys}$. See also Fig.~\ref{figS4} in Appendix~\ref{ap:FTLM}.)
The striking similarity between the eigenvalues of one-particle density matrices in Fig.~\ref{fig2}(a), obtained from the time propagation and from the FTLM, carries important information.
In general, agreement between the eigenvalues (i.e., the vanishing trace distance between the two matrices) implies the absence of any extensive set of one-body observables $\hat c_j^\dagger \hat c_l$ that would not thermalize. 
In addition, for our particular setup with a single electron, all multi-particle observables can be expressed by the one-particle operators.
The results in Fig.~\ref{fig2}(a) therefore suggest that all static properties of the electronic subsystem in the steady state are the same as in the Gibbs ensemble with temperature $T$. 
As a consequence, $T$ also becomes a well-defined temperature of the electronic subsystem.
A perfect agreement and strict thermalization is nevertheless expected only when the total system approaches the thermodynamic limit.

Our results for other parameters indicate that thermalization is visible as long as the system energy relative to the ground state considerably exceeds the phonon energy $\omega_0$.
An example where this is not fulfilled is shown in Fig.~\ref{fig2}(b) for $\omega_0/t_0=2.0$ and $E_{\rm sys} = 2t_0$ (here, the ground state energy is $E_{\rm gs} = - 2.596 t_0$).
In this case the system energy is located close to the bottom of many-body eigenenergy spectrum, where the density of states is low.
As a consequence, the time evolution is governed by strong temporal fluctuations. [See Fig.~\ref{fig2}(c) and also Fig.~\ref{fig_fss} of Appendix~\ref{sec:numerics}, which demonstrates that the temporal fluctuations are not an artifact of the numerical method].
Therefore, much longer time scales (but then also much larger systems) are needed to observe possible thermal behavior.
Note, however, that even for $\omega_0/t_0=2.0$ the difference between $\log  n_k(T)$ and $\log \bar n_k$  (obtained from the FTLM and from the time averaging, respectively) stays within $3 \sigma_k$, where $\sigma_k $ is the standard deviation characterizing time fluctuations of $\log n_k(t)$.

\section{Nonequilibrium dynamical correlation functions} \label{sec4}

Thermalization of eigenvalues of electronic one-particle density matrix implies thermalization of  all observables defined in terms of  fermionic  operators.
This, however, does not immediately extend to dynamical two-particle observables like the optical conductivity $\sigma(\omega)$.  The latter quantity is defined as a Fourier transform of the current-current correlation function  $\langle \hat \jmath(t) \hat \jmath(0) \rangle$~\cite{mahan2000}, with $\hat \jmath$ to be defined in Eq.~(\ref{def_current}).
Even if $\hat \jmath(0)$ is purely fermionic, the explicit form of $\hat \jmath(t)$ contains both fermionic and bosonic operators since the time evolution takes place under the Hamiltonian of the total system. Hence, $\sigma(\omega)$ cannot be obtained from the one-particle density matrix only. 
We therefore complement our study by calculating  the time-resolved optical conductivity $ \sigma(\omega,t)$~\cite{eckstein08,unterhinninghofen08,eckstein10,kanamori10,wall11,shimizu11,iyoda14,zala2014,lu15,shao15}.
In addition, we invoke a simple criterion to test whether dynamical correlation functions in the time-evolved pure state match the thermal form.

\subsection{Time-resolved optical conductivity} \label{tropcond}

The time-resolved optical conductivity provides access to the dynamical properties that are measured in pump-probe experiments~\cite{dalconte12,dalconte15}.
In addition, by means of a sum rule, it also provides access to a few static properties such as the electronic kinetic energy $E_{\rm kin}$, shown in Figs.~\ref{fig1} and~\ref{fig1b}.
Note that the optical sum rule
\begin{equation} \label{def_sumrule}
\int_{-\infty}^{\infty}d\omega \ \sigma'(\omega,t)=-\pi E_\mathrm{kin}(t),
\end{equation}
where $\sigma'(\omega,t) = \Re [\sigma(\omega,t)]$,
also holds far from equilibrium~\cite{zala2014}.
We calculate $ \sigma'(\omega,t)$ at arbitrary time $t$ after the quench without applying time-translation invariance~\cite{zala2014}, i.e., as a Fourier transform of the two-time correlation
\begin{equation} \label{sigtt}
\sigma(t',t)=-[E_\mathrm{kin}(t') +\int_{t}^{t'} dt'' \chi(t',t'') ],
\end{equation}
where the nonequilibrium dynamical susceptibility is defined as
\begin{equation} \label{chitt}
\chi(t',t'') =i \theta(t'-t'') \langle \psi_0 | [{\hat \jmath}(t'),{\hat \jmath}(t'')] | \psi_0 \rangle
\end{equation}
and $\hat \jmath$ denotes the current operator
\begin{equation} \label{def_current}
\hat \jmath= it_0 \sum_j ( e^{i \phi(t)} \hat c_{j+1}^\dagger \hat c_j - {\rm h.c.}).
\end{equation}
Details of the calculation are presented in Appendix~\ref{sp:sigom}.
We mimic the probe pulse of pump-probe experiments by a gaussian envelope with a width comparable to current experiments~\cite{dalconte15}.

\begin{figure}
\includegraphics[width=0.43\textwidth]{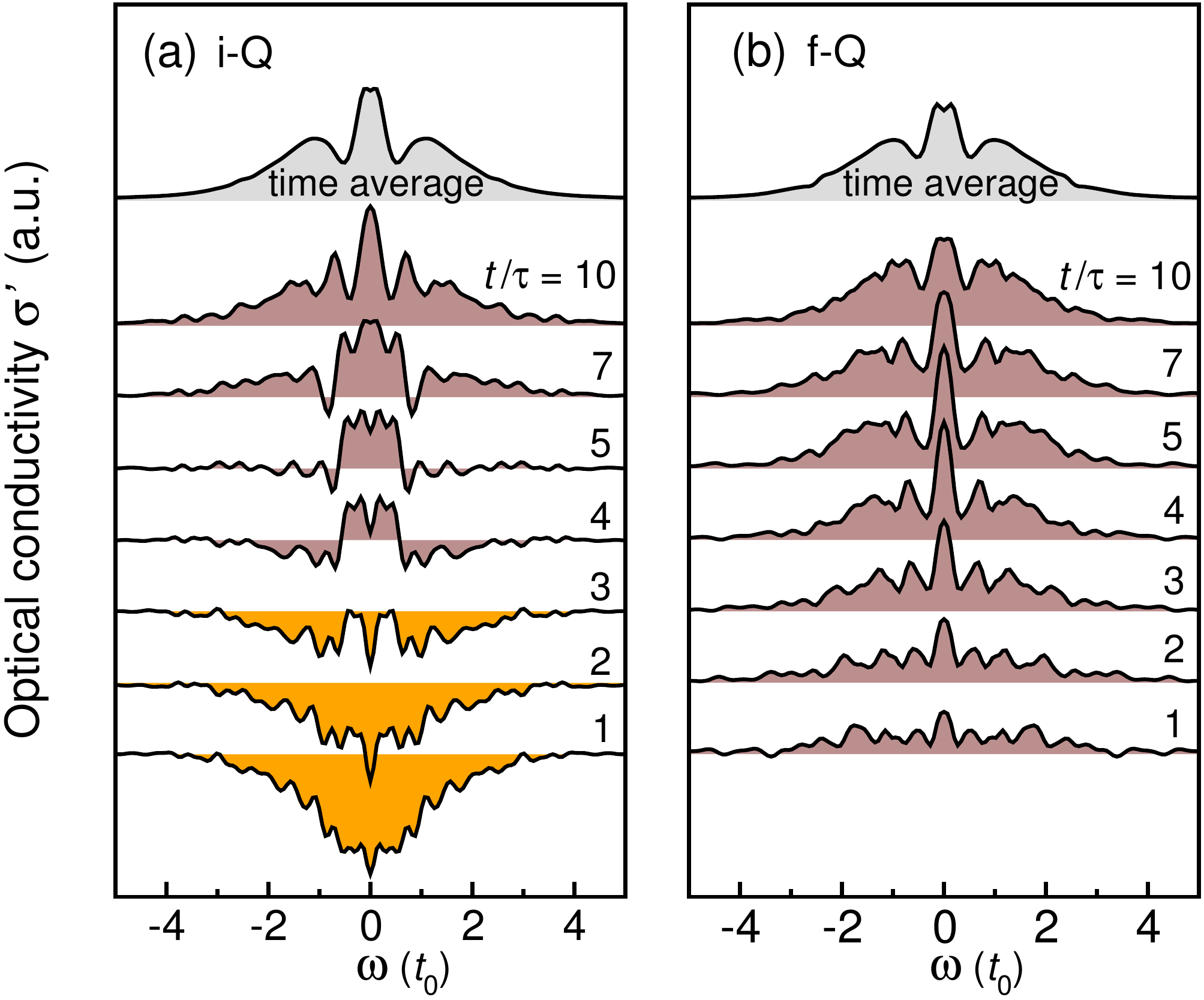}
\caption{
Time evolution of the optical conductivity $\sigma'(\omega,t)$ at the total energy $E_{\rm sys}=2t_0$
($\lambda=0.5$, $\omega_0/t_0=0.75$).
(a) Interaction quench (i-Q), (b) field quench (f-Q) with $F=1.0$.
The curves are offset vertically (from bottom to top) to display $\sigma'(\omega,t)$ at different times $t/\tau$.
The top curve in both panels is averaged in the time interval $t \in [50\tau,100\tau]$.
In panel (a), the sum rule of $\sigma'(\omega,t)$ is negative for $t \lesssim 4\tau$, consistent with Eq.~(\ref{def_sumrule}) and the results in Fig.~\ref{fig1}(a).
}
\label{fig3}
\end{figure}

Figure~\ref{fig3} displays the time evolution of $\sigma'(\omega,t)$ for quenches at the total energy $E_{\rm sys}=2t_0$.
For the interaction quench shown in Fig.~\ref{fig3}(a),
the states with positive single-particle energies $\epsilon_{k(t)} > 0$ are initially more occupied than the  low-energy states with $\epsilon_{k(t)} < 0$.
It leads to a positive  sign of the  kinetic  energy $E_\mathrm{kin}(t)=\sum_k \epsilon_{k(t)} n_{k}(t)$ [see Fig.~\ref{fig1}(a)] and hence to the negative optical sum rule $-\pi E_\mathrm{kin}(t)$. This results in negative values of the response function $\sigma'(\omega,t)$, which in turn represent the frequency range where  photoemission may occur.
At later times $\sigma'(\omega,t)$ becomes positive, it develops a well-defined peak around zero frequency corresponding to the Drude peak, and the side peaks reflecting multiple phonon excitations.
For the field quench at $F=1.0$ presented in Fig.~\ref{fig3}(b), $\sigma'(\omega,t)$ remains positive for all times, in strong contrast with the interaction quench discussed before.
Nevertheless, when entering the steady state at longer times, optical conductivities for different initial states display a very similar frequency dependence.

\subsection{Thermal form of dynamical correlations}

The results from Sec.~\ref{tropcond} give rise to an obvious question: to what extent does $\sigma'(\omega,t)$ resemble the linear response of a system in thermal equilibrium?
To answer this question, one could in principle follow the same procedure as in Sec.~\ref{sec3}, i.e., construct the corresponding Gibbs ensemble and compare the resulting thermal $\sigma'(\omega)$  with the time-resolved $\sigma'(\omega,t)$ in the steady state.
Instead, we show in the following that a simple manipulation of the dynamical correlations allows one to test whether their form is thermal without explicitly carrying out calculation in the Gibbs ensemble.
If the answer is affirmative, one can also extract the corresponding temperature of the Gibbs ensemble.

\begin{figure}
\includegraphics[width=0.48\textwidth]{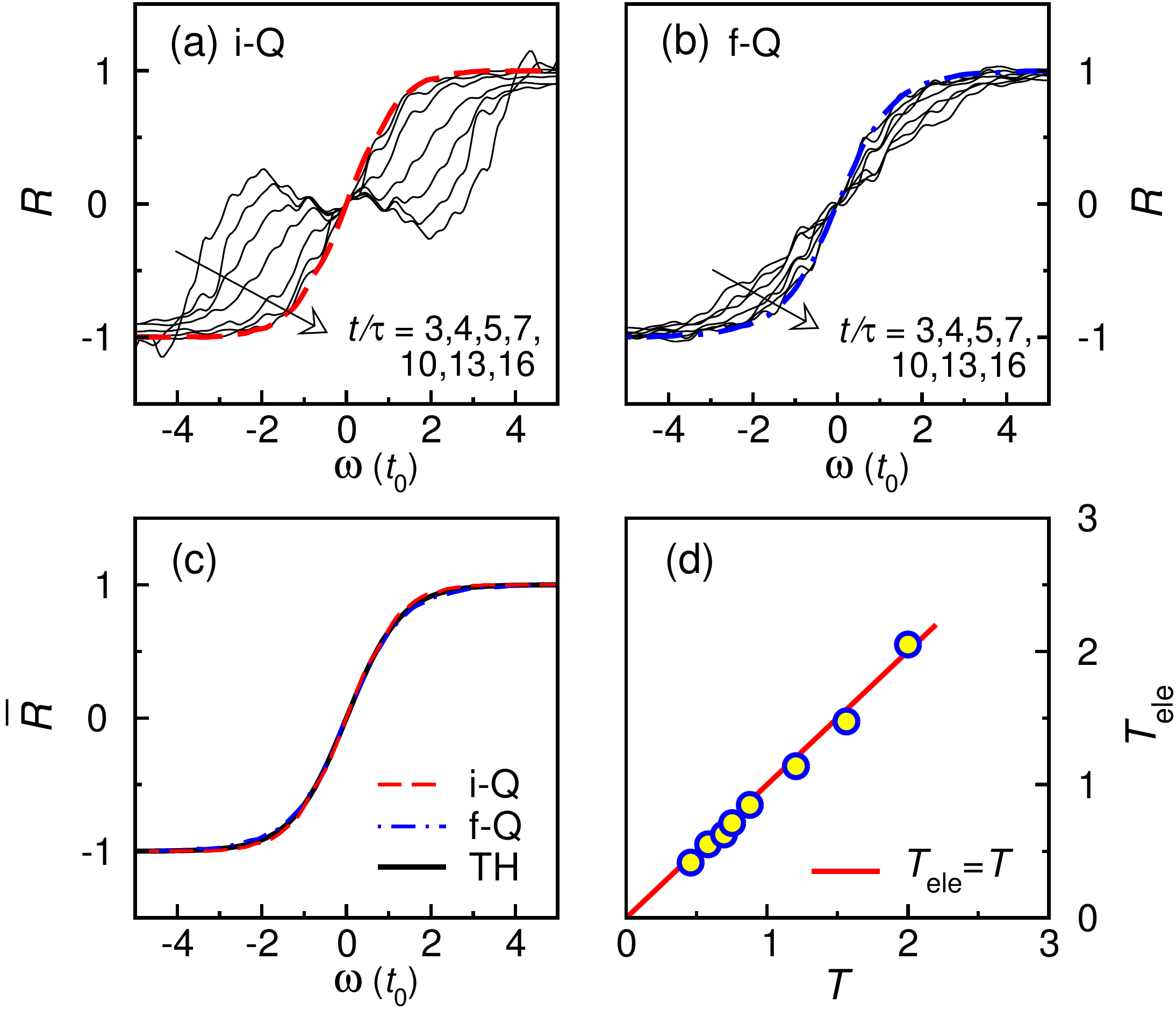}
\caption{
(a) and (b):
Time evolution of the dynamical correlation function $R(\omega,t)$ at the total energy $E_{\rm sys}=2t_0$ and $\lambda=0.5$, $\omega_0/t_0=0.75$.
Results are displayed at different times $t/\tau$
for (a) the interaction quench (i-Q) and (b) the field quench (f-Q) at $F=1.0$.
Thick dashed lines in both panels represent $\bar R(\omega) \equiv \langle R(\omega,t) \rangle_t$, where $\langle ... \rangle_t$ denotes time averaging in the interval $t \in [50\tau,100\tau]$.
Panel (c) compares $\bar R(\omega)$ with the thermal form $R(\omega)=\tanh[\omega/(2T)]$ (TH) at $T=0.68$.
Panel (d) shows comparison of the temperatures. 
$T_{\rm ele}$ is the temperature in the Gibbs ensemble obtained such that its electron kinetic energy matches the one in the steady state of the time-evolved state (as explained in Sec.~\ref{sec3}).
$T$ is the temperature, obtained by fitting $\bar R(\omega)$ with the thermal form $R(\omega)=\tanh[\omega/(2T)]$ from Eq.~(\ref{fom}).
The time evolution is in both cases performed after the field quench at $F = 1.0$.
}
\label{fig4}
\end{figure}

To demonstrate this idea we first note that in thermal equilibrium, the regular part of the optical conductivity can be expressed from the standard linear-response theory~\cite{mahan2000} as
\begin{equation}
\sigma'_\mathrm{reg}(\omega)={1-e^{-\omega/T}\over \omega}C(\omega), \label{sreg}
\end{equation}
where the dynamical current-current correlations are defined as
\begin{equation}
C(\omega) = \Re\int_0^\infty \mathrm{d}t e^{i\omega^+t} {\rm Tr}\{ \hat \rho_{\rm sys} \hat \jmath(t) \hat \jmath(0) \}. \label{com}
\end{equation}
As a central step, one can manipulate $C(\omega)$ to construct the function
\begin{equation}
R(\omega)={C(\omega)-C(-\omega)\over  C(\omega)+C(-\omega)}=\tanh \left({\omega\over 2T} \right), \label{fom}
\end{equation}
which allows for a direct extraction of the temperature.
Here, $C(\omega)$  is derived by using the Gibbs density matrix~(\ref{def_Gibbs})
[see Appendix~\ref{sp:com} for details of derivation].
We use the coefficient $R(\omega)=\tanh [{\omega/ (2T)} ]$ only as a measure for thermalization and do not explicitly calculate $C(\omega)$.

Our goal is to calculate a coefficient similar to the one in Eq.~(\ref{fom}), but calculated in the time-evolved wave function.
For this purpose, we define
\begin{equation}
R(\omega,t)={C(\omega,t)-C(-\omega,t)\over  C(\omega,t)+C(-\omega,t)},
\end{equation}
where the Fourier transform of the time-resolved current-current correlation is defined as
\begin{equation}
C(\omega,t) = \Re\int_0^\infty \mathrm{d}s e^{i\omega^+s}\langle \psi_0 | \hat \jmath(t+s) \hat \jmath(t) | \psi_0 \rangle \label{comt}.
\end{equation}
We calculate $C(\omega,t)$ numerically with the same method as the time-resolved optical conductivity $\sigma'(\omega,t)$ (see also Appendix~\ref{sp:sigom} for details).

Figures~\ref{fig4}(a) and \ref{fig4}(b) display $R(\omega,t)$ for two different initial states $| \psi_0 \rangle$ at equal total energy $E_{\rm sys}=2t_0$.
While the behavior of $R(\omega,t)$ at short times  clearly depends on the initial state, the differences are washed out at longer times.
In Fig.~\ref{fig4}(c) we compare the time-averaged $\langle R(\omega,t)\rangle_t$ in the steady state and show its independence of the initial state.
Remarkably, the form of $\langle R(\omega,t)\rangle_t$ perfectly agrees with the thermal functional form $\tanh [{\omega/ (2T)} ]$.

It is plausible, but by no means obvious that {\it (i)}  the temperature obtained from the Gibbs ensemble by matching the static one-particle electronic correlations
in the steady state (as done in Sec.~\ref{sec3}), and {\it (ii)} the temperature obtained from fitting the dynamical correlations $\langle R(\omega,t)\rangle_t$ 
with the thermal form $\tanh [{\omega/ (2T)} ]$, are equal.
Figure~\ref{fig4}(d) shows that such agreement indeed holds true.
It establishes an important link between the temperature, measured from the dynamical correlation functions, and the temperature of the electronic subsystem.

\section{Conclusion and discussion} \label{sec5}

We studied thermalization of a photoexcited electron coupled to quantum phonons using a state-of-the-art numerical method.
Even though in the simple Holstein model the relaxation channel is represented by a single (optical) phonon branch, many indicators for thermalization are fulfilled provided that the excitation energy is much larger than the phonon frequency.
First, we calculated all elements of the electronic one-particle density matrix, as well as the occupations of fermionic momenta, which are the corresponding eigenvalues.
We showed that in the steady state all those eigenvalues agree with the ones in the Gibbs ensemble, indicating thermalization of electronic correlations on the entire lattice.
This leads to the conclusion that the standard view of thermalization, which takes place only in a subsystem physically separated from the environment, is too restrictive.
Recently, a complete thermalization of one-particle observables towards predictions of the generalized Gibbs ensemble was reported for two integrable models~\cite{wright14,vidmar16}.
In contrast, the Holstein model is a generic (nonintegrable) system, and its eigenenergy spectrum is unbounded from above.
Our results hence give new insights into the nonequilibrium statistical mechanics of fermion-boson coupled systems and call for further studies, e.g., in the context of eigenstate thermalization~\cite{rigol08}.

Second, we have shown that the entire spectral distribution of the nonequilibrium dynamical current-current correlation function approaches in the steady state the thermal form.
The latter allows one to extract the temperature of the corresponding Gibbs ensemble.
Remarkably, the temperature obtained from the dynamical two-particle correlation function agrees with the one in the Gibbs ensemble that describes static one-particle electronic observables. 

From the point of view of time-resolved experiments, our model provides a reasonable description of the initial photocarrier relaxation through a single branch of bosonic excitations.
We denoted this process as the primary relaxation. 
Thermalization of electronic one-particle observables implies that the optical sum rule becomes, at a given excitation density, independent of the initial state.
In addition, the approach of the dynamical current-current correlation function to a simple thermal form, given by Eq.~(\ref{fom}),
reestablishes the potential of time-resolved optical spectroscopy~\cite{giannetti16} as a very efficient technique for studying thermalization (or the absence thereof) in solids. 

While our investigation focused on the primary relaxation regime in the context of pump-probe experiments, the current time resolution of many experiments
(e.g., the time-resolved ARPES~\cite{perfetti07,bovensiepen12,smallwood12b,piovera15,rameau15}) only allows measurements of  secondary relaxation processes.
Our results demonstrate high efficiency of the primary relaxation, leading to a thermal state even if the electron is strongly coupled to only a few bosonic  degrees of freedom.
Therefore,  during the secondary relaxation processes charge carriers  may be accurately described within the framework of  quasithermal evolution, i.e.,
by  a well-defined temperature  that evolves with time.
In this stage, electrons further exchange energy with other degrees of freedom of their environment and evolve towards the global equilibrium.

\begin{acknowledgments}
We acknowledge stimulating discussions with U. Bowensiepen, M. Eckstein, C. Giannetti and L. Perfetti. 
J.B. acknowledges discussions with A. Polkovnikov and M. Rigol as well as  the  support by the P1-0044 of ARRS, Slovenia.
L.V. acknowledges discussions with F. Heidrich-Meisner, F. Dorfner and E. Jeckelmann.
M.M. acknowledges support from the DEC-2013/11/B/ST3/00824 project of the Polish National Science Center.
This work was performed, in part, at the Center for Integrated Nanotechnologies, a U.S. Department of Energy, Office of Basic Energy Sciences user facility. 
\end{acknowledgments}

\appendix

\section{Numerical calculation within limited functional space}
\label{sec:numerics}

Here we provide more details about the limited functional space that we use in our Lanczos algorithm.
We truncate the Hilbert space by applying a limited functional space (LFS) generator~\cite{bonca99}.
We initiate the generation procedure by a starting state, which is a bare electron in a given momentum eigenstate $\hat c_k^{\dag} \ket{\emptyset}$.
We then apply the off-diagonal elements of $\hat H_{\rm sys}$ to this starting state.
The maximal number of generations of new states is controlled by the parameter $N_{\rm h}$.
We represent the entire set of states $ \left \{ \left \vert \phi_{k}^{(N_{\rm h})} \right \rangle \right  \}$ forming the LFS by a sum
\begin{align} \label{def_lfs}
\sum_{n_{\rm h}=0}^{N_{\rm h}}\left( \hat H_{\rm{kin}} + \hat H_{\rm{e-ph}} \right)^{n_{\rm h}} \hat c_k^{\dag} \ket{\emptyset}.
\end{align}
Further details about the method can be found, e.g., in Ref.~\cite{prelovvsek2013strongly}.
We set $N_{\rm h} = 22$ in our calculations and apply periodic boundary conditions on a lattice with $L=16$ sites.

The latter method, apart from being numerically exact for calculations of the Holstein polaron ground state~\cite{ku2002} and low-lying excited states~\cite{vidmar10}, can also be applied to study time-dependent problems~\cite{lev2011_1,golez2012,golez_prl2012,lev2015}.
By comparing it to other wavefunction-based methods, it has been recently shown to be the most efficient method to study the nonequilibrium dynamics of the Holstein polaron~\cite{lev2015} (see~\cite{brockt15} for a more recent methodological development, though).

\begin{figure}
\includegraphics[width=0.48\textwidth]{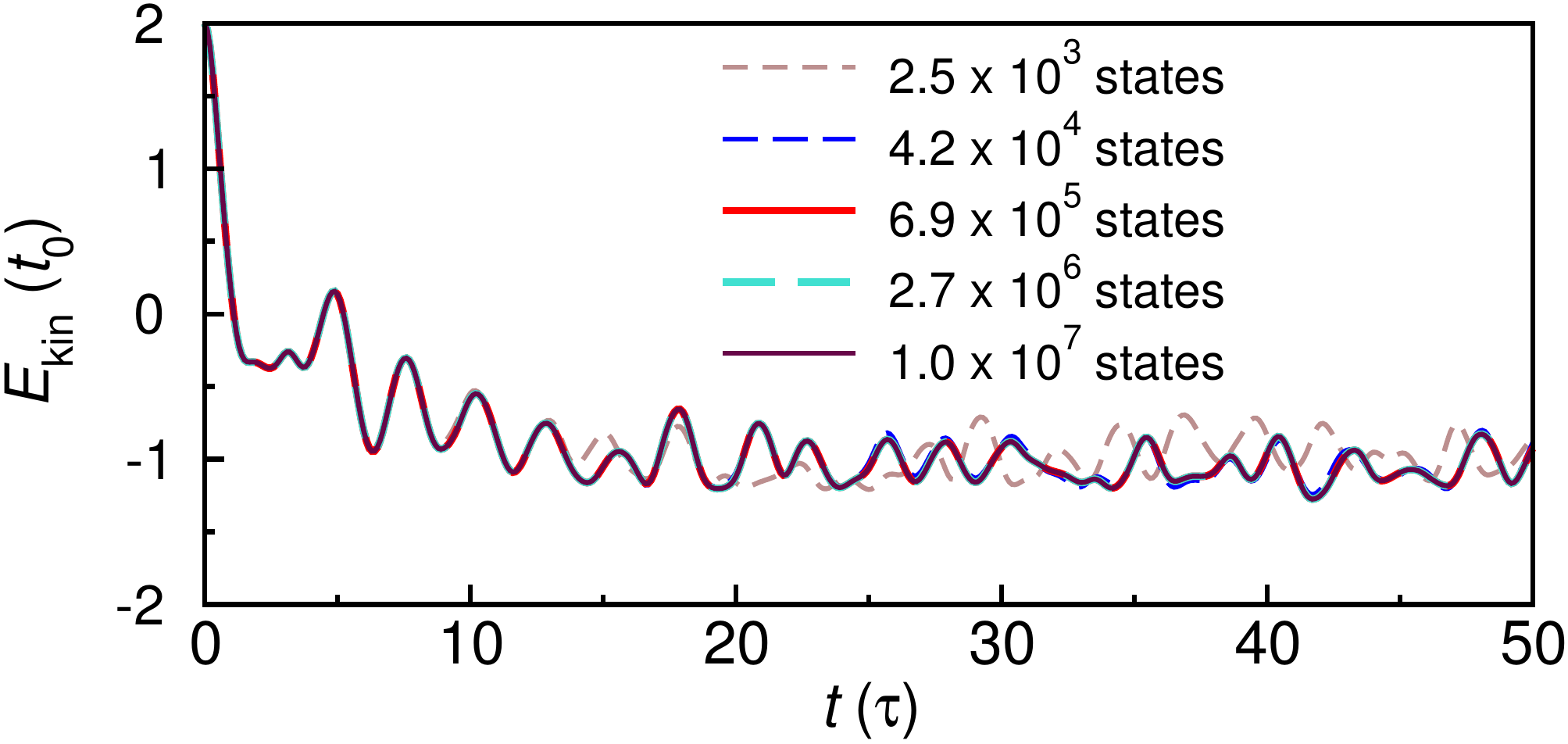}
\caption{
Time dependence of the electron kinetic energy  $E_{\mathrm{kin}}$ at $\lambda=0.5$ and $\omega_0/t_0=2.0$.
Different curves represent results for different sizes of the LFS.
We set $N_{\rm h}=10,14,18,20$ and $22$, which lead to
$2.5 \times 10^3$, $4.2 \times 10^4$, $6.9 \times 10^5$, $2.7 \times 10^6$ and $1.0 \times 10^7$ states in the LFS, respectively.
}
\label{fig_fss}
\end{figure}

We demonstrate the efficiency of calculation within the LFS in Fig.~\ref{fig_fss}.
We study the finite-size effects of the relaxation of electron kinetic energy $E_{\rm kin}$ at $\lambda=0.5$ and $\omega_0/t_0 = 2.0$.
The data for $N_{\rm h}=22$ are shown in Fig.~\ref{fig2}(c).
Since the system exhibits persistent oscillations around the steady-state value, one may wonder whether these are a finite-size effect or a real physical feature.
In Figure~\ref{fig_fss} we compare the data for five different sizes of the LFS obtained by setting $N_{\rm h}=10,14,18,20,22$ in Eq.~(\ref{def_lfs}).
Results show collapse of the data for all times of interest.
In particular, the minimal size of the LFS needed to get the converged results in this case only requires around $\sim 10^5$ states.

\section{Finite Temperature Lanczos Method} 
\label{ap:FTLM}
We briefly present the basic elements of the Finite Temperature Lanczos Method (FTLM).
We follow Refs.~\cite{jaklic00,prelovvsek2013strongly} where the method has been described in detail.
A straightforward calculation of the canonical thermodynamic average of an operator $\hat A$ in a finite system with $N_{st}$ states can be expressed
in an orthonormal basis $\{ |n \rangle \}$ as
\begin{equation}
\langle \hat A\rangle=\sum_{n=1}^{N_{st}}\langle n|e^{-\beta \hat H}\hat A|n\rangle
\biggm/
\sum_{n=1}^{N_{st}}\langle n|e^{-\beta \hat H}|n\rangle,
\label{Prelovsek:fh1}
\end{equation}
where $\beta = 1/T$ ($k_{\rm B}$ has been set to unity).
Using the high-temperature expansion of Eq.~(\ref{Prelovsek:fh1}) and the Lanczos procedure  with $M$ steps we obtain 
\begin{eqnarray}
\langle \hat A \rangle &=& Z^{-1}\sum_{n=1}^{N_{st}}\sum_{i=0}^M
e^{-\beta \epsilon^n_i}\langle n|\psi^n_i\rangle\langle\psi^n_i|\hat A|n
\rangle, \\
Z &=& \sum_{n=1}^{N_{st}}\sum_{i=0}^M e^{-\beta
\epsilon^n_i}\langle n|\psi^n_i\rangle\langle\psi^n_i|n
\rangle, \label{Prelovsek:fh4}
\end{eqnarray}
where $|\psi^n_i\rangle$ and $\epsilon^n_i$ are Lanczos functions and energies obtained from the starting state $\vert n\rangle$,
while the error of the approximation is $O(\beta^{M+1})$.
To allow computation of systems where $N_{st}\sim 10^7$ or more, the summation over $N_{st}$ states is replaced by a random sampling
over $\bar r$ random states $\vert r\rangle$
\begin{eqnarray}
\langle \hat A \rangle &=& \frac{N_{st}}{Z\bar r}\sum_{r=1}^{\bar r}\sum_{j=0}^M
e^{-\beta \epsilon^r_j}\langle r|\psi^r_j\rangle\langle\psi^r_j|\hat A|
r \rangle, \\
Z &=& \frac{N_{st}}{\bar r}\sum_{r=1}^{\bar r}\sum_{j=0}^M e^{-\beta
\epsilon^r_j}|\langle r|\psi^r_j\rangle|^2. \label{Prelovsek:fi1}
\end{eqnarray}
Random states $|r\rangle$ act as initial states for 
the  {Lanczos} iteration, resulting in $M$ eigenvalues 
$\epsilon^r_j$ with the corresponding $|\psi^r_j\rangle$.
Note that the summation over the random states runs over all $k$-sectors.
For each $k$-sector, we sample over $\bar r_k=20$ states. 

\begin{figure}
\includegraphics[width=0.48\textwidth]{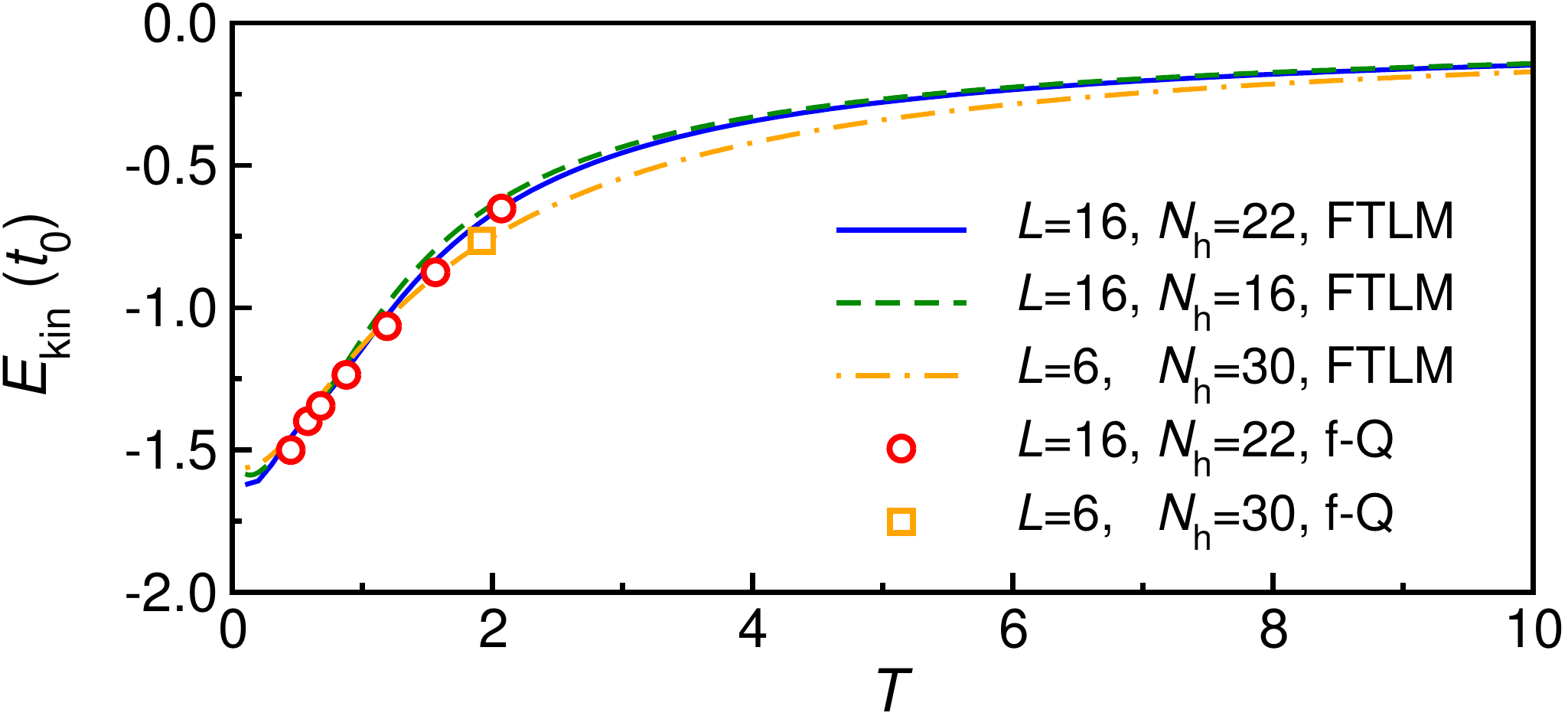}
\caption{
Lines:
Electron kinetic energy  $E_{\mathrm{kin}}$ in the Gibbs ensemble~(\ref{def_Gibbs}), versus the temperature $T$.
We use different lattice sizes $L$ and different sizes of the LFS~(\ref{def_lfs}), denoted by $N_{\rm h}$.
Symbols:
The long-time average of $E_{\mathrm{kin}}$ after the field quench at $F=1.0$.
The corresponding temperature is computed from the calculation of $\langle R(\omega,t) \rangle_t$ in the same system, and by fitting the latter quantity to the thermal form $\tanh[\omega/(2T)]$.
}
\label{figS4}
\end{figure}

In Fig.~\ref{figS4} we present $E_{\mathrm{kin}}(T)$ obtained by the FTLM and compare results for different lattice sizes $L=6$ and 16,
as well as different sizes of the LFS given by $N_{\mathrm h}=16$, 22 and 30.
At small $T$ deviations from the thermodynamic limit are expected predominantly due to the finite system size $L$,
while at larger $T$ when the electron coherence  length becomes shorter, deviations are expected due to a lack of adequate
number of phonon quanta contained in the LFS.
Despite  distinct lattice sizes chosen for comparison, $L=6$ and 16, variations of $E_{\mathrm{kin}}(T)$ are rather small throughout  the whole temperature range.
In addition, the time-propagation results (red circles) are mostly obtained in the temperature range of the smallest finite-size effects. 

\section{Dynamical current-current correlations in thermal equilibrium} 
\label{sp:com}

The Fourier transform of the dynamical current-current correlation $C(\omega)$ is defined as
\begin{eqnarray} \label{sup:C_init}
	C(\omega) &=& \Re\int_0^\infty \mathrm{d}t e^{i\omega^+t}\langle \hat \jmath(t) \hat \jmath(0) \rangle \\
	&=& \Re\int_0^\infty \mathrm{d}t e^{i\omega^+t} \mathrm{Tr}\{ \hat \rho_{\mathrm{sys}} \hat \jmath(t) \hat \jmath(0) \} \nonumber\\
	&=& \frac{1}{Z} \Re\int_0^\infty \mathrm{d}t e^{i\omega^+t} \mathrm{Tr}\{ e^{-\beta \hat H_{\mathrm{sys}}} e^{i \hat H_{\mathrm{sys}} t} \hat \jmath e^{-i \hat H_{\mathrm{sys}} t} \hat \jmath\}, \nonumber
\end{eqnarray}
where we used the Gibbs density matrix $\hat \rho_{\rm sys} = Z^{-1} e^{-\beta \hat H_{\rm sys}}$ and $Z = \mathrm{Tr}\{ e^{-\beta \hat H_{\rm sys}} \}$. 

Eigenstates of the Hamiltonian $\hat H_{\mathrm{sys}}|n\rangle = E_n |n\rangle$ form a complete set of states, $\sum_n |n \rangle \langle n| = 1$. Using the completeness relation we rewrite Eq.~(\ref{sup:C_init}) as
\begin{eqnarray}
	\label{sup:C_final}
	C(\omega) &=& \frac{1}{Z} \Re\int_0^\infty \mathrm{d}t e^{i\omega^+t} \times \nonumber \\ 
	&&\quad \quad \quad \quad \sum_{nm} e^{-\beta E_n} e^{i E_n t}\langle n| j |m\rangle e^{-i E_m t} \langle m| j |n\rangle \nonumber\\
	&=& \frac{1}{Z} \sum_{nm} e^{-\beta E_n} |\langle n| j |m\rangle|^2 2\pi \delta (E_n - E_m + \omega).
\end{eqnarray}
If we now set $\omega \rightarrow -\omega$, we can readily obtain
\begin{equation}
	C(-\omega) = e^{-\beta \omega} C(\omega),
\end{equation}
which we in turn use to produce Eq.~(\ref{fom}).

Note that the same result can be obtained using the fluctuation-dissipation theorem
\begin{equation}
\chi''(\omega) = -\frac{1}{2}(1-e^{-\beta\omega}) C(\omega),
\end{equation}
where $\chi''(\omega) = \Im [\chi(\omega)]$ represents the imaginary part of Fourier transform of the equilibrium dynamical susceptibility
\begin{equation}
\chi(t) =i \theta(t) \langle  [{\hat \jmath}(t),{\hat \jmath}(0)] \rangle.
\end{equation}
Equation~(\ref{fom}) can be reproduced using the symmetry $\chi''(-\omega) = - \chi''(\omega)$.



\section{Nonequilibrium optical conductivity}
\label{sp:sigom}

The evaluation of the two-time optical conductivity $\sigma(t',t)$, as defined in Eq.~(\ref{sigtt}), was implemented according to the procedure described in~\cite{zala2014}.
On the other hand, the implementation of a nonequilibrium dynamical susceptibility $\chi(t',t'')$ defined in Eq.~(\ref{chitt}) is a straightforward evaluation of the corresponding expectation value. All Fourier transformations, such as 
\begin{equation} 
	\sigma(\omega,t)=\int_0^{t_m} ds \ \sigma(t+s,t) e^{i\omega s},\label{sigomt}
\end{equation}
where $t_m$ denotes the maximum time of simulation, followed the definitions in~\cite{zala2014}.


Since our calculations are performed in a limited time window with $t_m = 100 \tau$, we use artificial spectral broadening to calculate $\sigma(\omega,t)$ and $C(\omega,t)$, smoothing the non-physical artifacts.
The applied broadening replaces $e^{i\omega s}$ in Eq.~(\ref{sigomt}) with $e^{i\omega s}e^{-{s^2}/{2W^2}}$, where we set the width of the Gaussian envelope to $W=10 \tau$.

\bibliographystyle{biblev1}
\bibliography{Polarons}

\begin{thebibliography}{10}
\expandafter\ifx\csname url\endcsname\relax
  \def\url#1{{\tt #1}}\fi
\expandafter\ifx\csname urlprefix\endcsname\relax\def\urlprefix{URL }\fi
\expandafter\ifx\csname bibinfo\endcsname\relax\def\bibinfo#1#2{#2}\fi
\expandafter\ifx\csname eprint\endcsname\relax\def\eprint#1{\url{#1}}\fi

\bibitem{polkovnikov11}
\bibinfo{author}{A.~Polkovnikov}, \bibinfo{author}{K.~Sengupta},
  \bibinfo{author}{A.~Silva}, and \bibinfo{author}{M.~Vengalattore},
  \bibinfo{title}{Nonequilibrium dynamics of closed interacting quantum
  systems},
  \bibinfo{journal}{\href{http://dx.doi.org/10.1103/RevModPhys.83.863}{Rev.
  Mod. Phys}} \href{http://dx.doi.org/10.1103/RevModPhys.83.863}{{\bf
  \bibinfo{volume}{83}}, \bibinfo{pages}{863}}
  (\href{http://dx.doi.org/10.1103/RevModPhys.83.863}{\bibinfo{year}{2011}}).

\bibitem{eisert15}
\bibinfo{author}{C.~Gogolin} and \bibinfo{author}{J.~Eisert},
  \bibinfo{title}{Equilibration, thermalisation, and the emergence of
  statistical mechanics in closed quantum systems},
  \bibinfo{journal}{\href{http://dx.doi.org/10.1088/0034-4885/79/5/056001}{Rep.
  Prog. Phys.}} \href{http://dx.doi.org/10.1088/0034-4885/79/5/056001}{{\bf
  \bibinfo{volume}{79}}, \bibinfo{pages}{056001}}
  (\href{http://dx.doi.org/10.1088/0034-4885/79/5/056001}{\bibinfo{year}{2016}}).

\bibitem{dalessio_kafri_15}
\bibinfo{author}{L.~D'Alessio}, \bibinfo{author}{Y.~Kafri},
  \bibinfo{author}{A.~Polkovnikov}, and \bibinfo{author}{M.~Rigol},
  \bibinfo{title}{From quantum chaos and eigenstate thermalization to
  statistical mechanics and thermodynamics},
  \href{http://arxiv.org/abs/1509.06411}{\bibinfo{howpublished}{arXiv:1509.06411}}.

\bibitem{rigol08}
\bibinfo{author}{M.~Rigol}, \bibinfo{author}{V.~Dunjko}, and
  \bibinfo{author}{M.~Olshanii}, \bibinfo{title}{Thermalization and its
  mechanism for generic isolated quantum systems},
  \bibinfo{journal}{\href{http://dx.doi.org/10.1038/nature06838}{Nature
  (London)}} \href{http://dx.doi.org/10.1038/nature06838}{{\bf
  \bibinfo{volume}{452}}, \bibinfo{pages}{854}}
  (\href{http://dx.doi.org/10.1038/nature06838}{\bibinfo{year}{2008}}).

\bibitem{Linden2009}
\bibinfo{author}{N.~Linden}, \bibinfo{author}{S.~Popescu},
  \bibinfo{author}{A.~J. Short}, and \bibinfo{author}{A.~Winter},
  \bibinfo{title}{Quantum mechanical evolution towards thermal equilibrium},
  \bibinfo{journal}{\href{http://dx.doi.org/10.1103/PhysRevE.79.061103}{Phys.
  Rev. E}} \href{http://dx.doi.org/10.1103/PhysRevE.79.061103}{{\bf
  \bibinfo{volume}{79}}, \bibinfo{pages}{061103}}
  (\href{http://dx.doi.org/10.1103/PhysRevE.79.061103}{\bibinfo{year}{2009}}).

\bibitem{giannetti16}
\bibinfo{author}{C.~Giannetti}, \bibinfo{author}{M.~Capone},
  \bibinfo{author}{D.~Fausti}, \bibinfo{author}{M.~Fabrizio},
  \bibinfo{author}{F.~Parmigiani}, and \bibinfo{author}{D.~Mihailovic},
  \bibinfo{title}{Ultrafast optical spectroscopy of strongly correlated
  materials and high-temperature superconductors: a non-equilibrium approach},
  \href{http://arxiv.org/abs/1601.07204}{\bibinfo{howpublished}{arXiv:1601.07204}}.

\bibitem{dalconte15}
\bibinfo{author}{S.~Dal~Conte}, \bibinfo{author}{L.~Vidmar},
  \bibinfo{author}{D.~Gole\v{z}}, \bibinfo{author}{M.~Mierzejewski},
  \bibinfo{author}{G.~Soavi}, \bibinfo{author}{S.~Peli},
  \bibinfo{author}{F.~Banfi}, \bibinfo{author}{G.~Ferrini},
  \bibinfo{author}{R.~Comin}, \bibinfo{author}{B.~M. Ludbrook},
  \bibinfo{author}{L.~Chauviere}, \bibinfo{author}{N.~D. Zhigadlo},
  \bibinfo{author}{H.~Eisaki}, \bibinfo{author}{M.~Greven},
  \bibinfo{author}{S.~Lupi}, \bibinfo{author}{A.~Damascelli},
  \bibinfo{author}{D.~Brida}, \bibinfo{author}{M.~Capone},
  \bibinfo{author}{J.~Bon\v{c}a}, \bibinfo{author}{G.~Cerullo}, and
  \bibinfo{author}{C.~Giannetti}, \bibinfo{title}{Snapshots of the retarded
  interaction of charge carriers with ultrafast fluctuations in cuprates},
  \bibinfo{journal}{\href{http://dx.doi.org/10.1038/nphys3265}{Nature Physics}}
  \href{http://dx.doi.org/10.1038/nphys3265}{{\bf \bibinfo{volume}{11}},
  \bibinfo{pages}{421}}
  (\href{http://dx.doi.org/10.1038/nphys3265}{\bibinfo{year}{2015}}).

\bibitem{okamoto2010}
\bibinfo{author}{H.~Okamoto}, \bibinfo{author}{T.~Miyagoe},
  \bibinfo{author}{K.~Kobayashi}, \bibinfo{author}{H.~Uemura},
  \bibinfo{author}{H.~Nishioka}, \bibinfo{author}{H.~Matsuzaki},
  \bibinfo{author}{A.~Sawa}, and \bibinfo{author}{Y.~Tokura},
  \bibinfo{title}{Ultrafast charge dynamics in photoexcited
  $\mathrm{Nd}_{2}\mathrm{CuO}_{4}$ and $\mathrm{La}_{2}\mathrm{CuO}_{4}$
  cuprate compounds investigated by femtosecond absorption spectroscopy},
  \bibinfo{journal}{\href{http://dx.doi.org/10.1103/PhysRevB.82.060513}{Phys.
  Rev. B}} \href{http://dx.doi.org/10.1103/PhysRevB.82.060513}{{\bf
  \bibinfo{volume}{82}}, \bibinfo{pages}{060513}}
  (\href{http://dx.doi.org/10.1103/PhysRevB.82.060513}{\bibinfo{year}{2010}}).

\bibitem{gadermaier10}
\bibinfo{author}{C.~Gadermaier}, \bibinfo{author}{A.~S. Alexandrov},
  \bibinfo{author}{V.~V. Kabanov}, \bibinfo{author}{P.~Kusar},
  \bibinfo{author}{T.~Mertelj}, \bibinfo{author}{X.~Yao},
  \bibinfo{author}{C.~Manzoni}, \bibinfo{author}{D.~Brida},
  \bibinfo{author}{G.~Cerullo}, and \bibinfo{author}{D.~Mihailovic},
  \bibinfo{title}{Electron-phonon coupling in high-temperature cuprate
  superconductors determined from electron relaxation rates},
  \bibinfo{journal}{\href{http://dx.doi.org/10.1103/PhysRevLett.105.257001}{Phys.
  Rev. Lett.}} \href{http://dx.doi.org/10.1103/PhysRevLett.105.257001}{{\bf
  \bibinfo{volume}{105}}, \bibinfo{pages}{257001}}
  (\href{http://dx.doi.org/10.1103/PhysRevLett.105.257001}{\bibinfo{year}{2010}}).

\bibitem{dalconte12}
\bibinfo{author}{S.~Dal~Conte}, \bibinfo{author}{C.~Giannetti},
  \bibinfo{author}{G.~Coslovich}, \bibinfo{author}{F.~Cilento},
  \bibinfo{author}{D.~Bossini}, \bibinfo{author}{T.~Abebaw},
  \bibinfo{author}{F.~Banfi}, \bibinfo{author}{G.~Ferrini},
  \bibinfo{author}{H.~Eisaki}, \bibinfo{author}{M.~Greven},
  \bibinfo{author}{A.~Damascelli}, \bibinfo{author}{D.~van~der Marel}, and
  \bibinfo{author}{F.~Parmigiani}, \bibinfo{title}{Disentangling the electronic
  and phononic glue in a high-$\mathrm{T}$c superconductor},
  \bibinfo{journal}{\href{http://dx.doi.org/10.1126/science.1216765}{Science}}
  \href{http://dx.doi.org/10.1126/science.1216765}{{\bf \bibinfo{volume}{335}},
  \bibinfo{pages}{1600}}
  (\href{http://dx.doi.org/10.1126/science.1216765}{\bibinfo{year}{2012}}).

\bibitem{gadermaier14}
\bibinfo{author}{C.~Gadermaier}, \bibinfo{author}{V.~V. Kabanov},
  \bibinfo{author}{A.~S. Alexandrov}, \bibinfo{author}{L.~Stojchevska},
  \bibinfo{author}{T.~Mertelj}, \bibinfo{author}{C.~Manzoni},
  \bibinfo{author}{G.~Cerullo}, \bibinfo{author}{N.~D. Zhigadlo},
  \bibinfo{author}{J.~Karpinski}, \bibinfo{author}{Y.~Q. Cai},
  \bibinfo{author}{X.~Yao}, \bibinfo{author}{Y.~Toda},
  \bibinfo{author}{M.~Oda}, \bibinfo{author}{S.~Sugai}, and
  \bibinfo{author}{D.~Mihailovic}, \bibinfo{title}{Strain-induced enhancement
  of the electron energy relaxation in strongly correlated superconductors},
  \bibinfo{journal}{\href{http://dx.doi.org/10.1103/PhysRevX.4.011056}{Phys.
  Rev. X}} \href{http://dx.doi.org/10.1103/PhysRevX.4.011056}{{\bf
  \bibinfo{volume}{4}}, \bibinfo{pages}{011056}}
  (\href{http://dx.doi.org/10.1103/PhysRevX.4.011056}{\bibinfo{year}{2014}}).

\bibitem{novelli14}
\bibinfo{author}{F.~Novelli}, \bibinfo{author}{G.~De~Filippis},
  \bibinfo{author}{V.~Cataudella}, \bibinfo{author}{M.~Esposito},
  \bibinfo{author}{I.~Vergara}, \bibinfo{author}{F.~Cilento},
  \bibinfo{author}{E.~Sindici}, \bibinfo{author}{A.~Amaricci},
  \bibinfo{author}{C.~Giannetti}, \bibinfo{author}{D.~Prabhakaran},
  \bibinfo{author}{S.~Wall}, \bibinfo{author}{A.~Perucchi},
  \bibinfo{author}{S.~Dal~Conte}, \bibinfo{author}{G.~Cerullo},
  \bibinfo{author}{M.~Capone}, \bibinfo{author}{A.~Mishchenko},
  \bibinfo{author}{M.~Gr{\"u}ninger}, \bibinfo{author}{N.~Nagaosa},
  \bibinfo{author}{F.~Parmigiani}, and \bibinfo{author}{D.~Fausti},
  \bibinfo{title}{Witnessing the formation and relaxation of dressed
  quasi-particles in a strongly correlated electron system},
  \bibinfo{journal}{\href{http://dx.doi.org/10.1038/ncomms6112}{Nature
  Communications}} \href{http://dx.doi.org/10.1038/ncomms6112}{{\bf
  \bibinfo{volume}{5}}, \bibinfo{pages}{5112}}
  (\href{http://dx.doi.org/10.1038/ncomms6112}{\bibinfo{year}{2014}}).

\bibitem{kaganov1957}
\bibinfo{author}{M.~Kaganov}, \bibinfo{author}{I.~Lifshitz}, and
  \bibinfo{author}{L.~Tanatarov}, \bibinfo{title}{Relaxation between electrons
  and the crystalline lattice}, \bibinfo{journal}{Sov. Phys. JETP} {\bf
  \bibinfo{volume}{4}}, \bibinfo{pages}{173}  (\bibinfo{year}{1957}).

\bibitem{allen87}
\bibinfo{author}{P.~B. Allen}, \bibinfo{title}{Theory of thermal relaxation of
  electrons in metals},
  \bibinfo{journal}{\href{http://dx.doi.org/10.1103/PhysRevLett.59.1460}{Phys.
  Rev. Lett.}} \href{http://dx.doi.org/10.1103/PhysRevLett.59.1460}{{\bf
  \bibinfo{volume}{59}}, \bibinfo{pages}{1460}}
  (\href{http://dx.doi.org/10.1103/PhysRevLett.59.1460}{\bibinfo{year}{1987}}).

\bibitem{golez_prl2012}
\bibinfo{author}{D.~Gole\ifmmode~\check{z}\else \v{z}\fi{}},
  \bibinfo{author}{J.~Bon\ifmmode~\check{c}\else \v{c}\fi{}a},
  \bibinfo{author}{L.~Vidmar}, and \bibinfo{author}{S.~A. Trugman},
  \bibinfo{title}{{Relaxation dynamics of the Holstein polaron}},
  \bibinfo{journal}{\href{http://dx.doi.org/10.1103/PhysRevLett.109.236402}{Phys.
  Rev. Lett.}} \href{http://dx.doi.org/10.1103/PhysRevLett.109.236402}{{\bf
  \bibinfo{volume}{109}}, \bibinfo{pages}{236402}}
  (\href{http://dx.doi.org/10.1103/PhysRevLett.109.236402}{\bibinfo{year}{2012}}).

\bibitem{defilippis12}
\bibinfo{author}{G.~De~Filippis}, \bibinfo{author}{V.~Cataudella},
  \bibinfo{author}{E.~A. Nowadnick}, \bibinfo{author}{T.~P. Devereaux},
  \bibinfo{author}{A.~S. Mishchenko}, and \bibinfo{author}{N.~Nagaosa},
  \bibinfo{title}{{Quantum dynamics of the Hubbard-Holstein model in
  equilibrium and nonequilibrium: Application to pump-probe phenomena}},
  \bibinfo{journal}{\href{http://dx.doi.org/10.1103/PhysRevLett.109.176402}{Phys.
  Rev. Lett.}} \href{http://dx.doi.org/10.1103/PhysRevLett.109.176402}{{\bf
  \bibinfo{volume}{109}}, \bibinfo{pages}{176402}}
  (\href{http://dx.doi.org/10.1103/PhysRevLett.109.176402}{\bibinfo{year}{2012}}).

\bibitem{matsueda12}
\bibinfo{author}{H.~Matsueda}, \bibinfo{author}{S.~Sota},
  \bibinfo{author}{T.~Tohyama}, and \bibinfo{author}{S.~Maekawa},
  \bibinfo{title}{Relaxation dynamics of photocarriers in one-dimensional
  $\mathrm{M}$ott insulators coupled to phonons},
  \bibinfo{journal}{\href{http://dx.doi.org/10.1143/JPSJ.81.013701}{J. Phys.
  Soc. Jpn.}} \href{http://dx.doi.org/10.1143/JPSJ.81.013701}{{\bf
  \bibinfo{volume}{81}}, \bibinfo{pages}{013701}}
  (\href{http://dx.doi.org/10.1143/JPSJ.81.013701}{\bibinfo{year}{2012}}).

\bibitem{kennes10}
\bibinfo{author}{D.~M. Kennes} and \bibinfo{author}{V.~Meden},
  \bibinfo{title}{Relaxation dynamics of an exactly solvable electron-phonon
  model},
  \bibinfo{journal}{\href{http://dx.doi.org/10.1103/PhysRevB.82.085109}{Phys.
  Rev. B}} \href{http://dx.doi.org/10.1103/PhysRevB.82.085109}{{\bf
  \bibinfo{volume}{82}}, \bibinfo{pages}{085109}}
  (\href{http://dx.doi.org/10.1103/PhysRevB.82.085109}{\bibinfo{year}{2010}}).

\bibitem{kemper13}
\bibinfo{author}{A.~F. Kemper}, \bibinfo{author}{M.~Sentef},
  \bibinfo{author}{B.~Moritz}, \bibinfo{author}{C.~C. Kao},
  \bibinfo{author}{Z.~X. Shen}, \bibinfo{author}{J.~K. Freericks}, and
  \bibinfo{author}{T.~P. Devereaux}, \bibinfo{title}{Mapping of unoccupied
  states and relevant bosonic modes via the time-dependent momentum
  distribution},
  \bibinfo{journal}{\href{http://dx.doi.org/10.1103/PhysRevB.87.235139}{Phys.
  Rev. B}} \href{http://dx.doi.org/10.1103/PhysRevB.87.235139}{{\bf
  \bibinfo{volume}{87}}, \bibinfo{pages}{235139}}
  (\href{http://dx.doi.org/10.1103/PhysRevB.87.235139}{\bibinfo{year}{2013}}).

\bibitem{sentef13}
\bibinfo{author}{M.~Sentef}, \bibinfo{author}{A.~F. Kemper},
  \bibinfo{author}{B.~Moritz}, \bibinfo{author}{J.~K. Freericks},
  \bibinfo{author}{Z.-X. Shen}, and \bibinfo{author}{T.~P. Devereaux},
  \bibinfo{title}{Examining electron-boson coupling using time-resolved
  spectroscopy},
  \bibinfo{journal}{\href{http://dx.doi.org/10.1103/PhysRevX.3.041033}{Phys.
  Rev. X}} \href{http://dx.doi.org/10.1103/PhysRevX.3.041033}{{\bf
  \bibinfo{volume}{3}}, \bibinfo{pages}{041033}}
  (\href{http://dx.doi.org/10.1103/PhysRevX.3.041033}{\bibinfo{year}{2013}}).

\bibitem{werner2013}
\bibinfo{author}{P.~Werner} and \bibinfo{author}{M.~Eckstein},
  \bibinfo{title}{{Phonon-enhanced relaxation and excitation in the
  Holstein-Hubbard model}},
  \bibinfo{journal}{\href{http://dx.doi.org/10.1103/PhysRevB.88.165108}{Phys.
  Rev. B}} \href{http://dx.doi.org/10.1103/PhysRevB.88.165108}{{\bf
  \bibinfo{volume}{88}}, \bibinfo{pages}{165108}}
  (\href{http://dx.doi.org/10.1103/PhysRevB.88.165108}{\bibinfo{year}{2013}}).

\bibitem{hohenadler13}
\bibinfo{author}{M.~Hohenadler}, \bibinfo{title}{{Charge and spin correlations
  of a Peierls insulator after a quench}},
  \bibinfo{journal}{\href{http://dx.doi.org/10.1103/PhysRevB.88.064303}{Phys.
  Rev. B}} \href{http://dx.doi.org/10.1103/PhysRevB.88.064303}{{\bf
  \bibinfo{volume}{88}}, \bibinfo{pages}{064303}}
  (\href{http://dx.doi.org/10.1103/PhysRevB.88.064303}{\bibinfo{year}{2013}}).

\bibitem{baranov14}
\bibinfo{author}{V.~V. Baranov} and \bibinfo{author}{V.~V. Kabanov},
  \bibinfo{title}{Theory of electronic relaxation in a metal excited by an
  ultrashort optical pump},
  \bibinfo{journal}{\href{http://dx.doi.org/10.1103/PhysRevB.89.125102}{Phys.
  Rev. B}} \href{http://dx.doi.org/10.1103/PhysRevB.89.125102}{{\bf
  \bibinfo{volume}{89}}, \bibinfo{pages}{125102}}
  (\href{http://dx.doi.org/10.1103/PhysRevB.89.125102}{\bibinfo{year}{2014}}).

\bibitem{kemper14}
\bibinfo{author}{A.~F. Kemper}, \bibinfo{author}{M.~A. Sentef},
  \bibinfo{author}{B.~Moritz}, \bibinfo{author}{J.~K. Freericks}, and
  \bibinfo{author}{T.~P. Devereaux}, \bibinfo{title}{Effect of dynamical
  spectral weight redistribution on effective interactions in time-resolved
  spectroscopy},
  \bibinfo{journal}{\href{http://dx.doi.org/10.1103/PhysRevB.90.075126}{Phys.
  Rev. B}} \href{http://dx.doi.org/10.1103/PhysRevB.90.075126}{{\bf
  \bibinfo{volume}{90}}, \bibinfo{pages}{075126}}
  (\href{http://dx.doi.org/10.1103/PhysRevB.90.075126}{\bibinfo{year}{2014}}).

\bibitem{werner15}
\bibinfo{author}{P.~Werner} and \bibinfo{author}{M.~Eckstein},
  \bibinfo{title}{{Field-induced polaron formation in the Holstein-Hubbard
  model}},
  \bibinfo{journal}{\href{http://dx.doi.org/10.1209/0295-5075/109/37002}{EPL
  (Europhysics Letters)}}
  \href{http://dx.doi.org/10.1209/0295-5075/109/37002}{{\bf
  \bibinfo{volume}{109}}, \bibinfo{pages}{37002}}
  (\href{http://dx.doi.org/10.1209/0295-5075/109/37002}{\bibinfo{year}{2015}}).

\bibitem{aoki2015}
\bibinfo{author}{Y.~Murakami}, \bibinfo{author}{P.~Werner},
  \bibinfo{author}{N.~Tsuji}, and \bibinfo{author}{H.~Aoki},
  \bibinfo{title}{{Interaction quench in the Holstein model: Thermalization
  crossover from electron- to phonon-dominated relaxation}},
  \bibinfo{journal}{\href{http://dx.doi.org/10.1103/PhysRevB.91.045128}{Phys.
  Rev. B}} \href{http://dx.doi.org/10.1103/PhysRevB.91.045128}{{\bf
  \bibinfo{volume}{91}}, \bibinfo{pages}{045128}}
  (\href{http://dx.doi.org/10.1103/PhysRevB.91.045128}{\bibinfo{year}{2015}}).

\bibitem{lev2015}
\bibinfo{author}{F.~Dorfner}, \bibinfo{author}{L.~Vidmar},
  \bibinfo{author}{C.~Brockt}, \bibinfo{author}{E.~Jeckelmann}, and
  \bibinfo{author}{F.~Heidrich-Meisner}, \bibinfo{title}{{Real-time decay of a
  highly excited charge carrier in the one-dimensional Holstein model}},
  \bibinfo{journal}{\href{http://dx.doi.org/10.1103/PhysRevB.91.104302}{Phys.
  Rev. B}} \href{http://dx.doi.org/10.1103/PhysRevB.91.104302}{{\bf
  \bibinfo{volume}{91}}, \bibinfo{pages}{104302}}
  (\href{http://dx.doi.org/10.1103/PhysRevB.91.104302}{\bibinfo{year}{2015}}).

\bibitem{sayyad15}
\bibinfo{author}{S.~Sayyad} and \bibinfo{author}{M.~Eckstein},
  \bibinfo{title}{Coexistence of excited polarons and metastable delocalized
  states in photoinduced metals},
  \bibinfo{journal}{\href{http://dx.doi.org/10.1103/PhysRevB.91.104301}{Phys.
  Rev. B}} \href{http://dx.doi.org/10.1103/PhysRevB.91.104301}{{\bf
  \bibinfo{volume}{91}}, \bibinfo{pages}{104301}}
  (\href{http://dx.doi.org/10.1103/PhysRevB.91.104301}{\bibinfo{year}{2015}}).

\bibitem{mishchenko15}
\bibinfo{author}{A.~S. Mishchenko}, \bibinfo{author}{N.~Nagaosa},
  \bibinfo{author}{G.~De~Filippis}, \bibinfo{author}{A.~de~Candia}, and
  \bibinfo{author}{V.~Cataudella}, \bibinfo{title}{Mobility of {H}olstein
  polaron at finite temperature: An unbiased approach},
  \bibinfo{journal}{\href{http://dx.doi.org/10.1103/PhysRevLett.114.146401}{Phys.
  Rev. Lett.}} \href{http://dx.doi.org/10.1103/PhysRevLett.114.146401}{{\bf
  \bibinfo{volume}{114}}, \bibinfo{pages}{146401}}
  (\href{http://dx.doi.org/10.1103/PhysRevLett.114.146401}{\bibinfo{year}{2015}}).

\bibitem{werner16}
\bibinfo{author}{P.~Werner} and \bibinfo{author}{M.~Eckstein},
  \bibinfo{title}{{Effective doublon and hole temperatures in the photo-doped
  dynamic Hubbard model}},
  \bibinfo{journal}{\href{http://dx.doi.org/10.1063/1.4935245}{Struct. Dyn.}}
  \href{http://dx.doi.org/10.1063/1.4935245}{{\bf \bibinfo{volume}{3}},
  \bibinfo{pages}{023603}}
  (\href{http://dx.doi.org/10.1063/1.4935245}{\bibinfo{year}{2016}}).

\bibitem{rizzi15}
\bibinfo{author}{V.~Rizzi}, \bibinfo{author}{T.~N. Todorov},
  \bibinfo{author}{J.~J. Kohanoff}, and \bibinfo{author}{A.~A. Correa},
  \bibinfo{title}{Electron-phonon thermalization in a scalable method for
  real-time quantum dynamics},
  \bibinfo{journal}{\href{http://dx.doi.org/10.1103/PhysRevB.93.024306}{Phys.
  Rev. B}} \href{http://dx.doi.org/10.1103/PhysRevB.93.024306}{{\bf
  \bibinfo{volume}{93}}, \bibinfo{pages}{024306}}
  (\href{http://dx.doi.org/10.1103/PhysRevB.93.024306}{\bibinfo{year}{2016}}).

\bibitem{golez2014}
\bibinfo{author}{D.~Gole\ifmmode~\check{z}\else \v{z}\fi{}},
  \bibinfo{author}{J.~Bon\ifmmode~\check{c}\else \v{c}\fi{}a},
  \bibinfo{author}{M.~Mierzejewski}, and \bibinfo{author}{L.~Vidmar},
  \bibinfo{title}{Mechanism of ultrafast relaxation of a photo-carrier in
  antiferromagnetic spin background},
  \bibinfo{journal}{\href{http://dx.doi.org/10.1103/PhysRevB.89.165118}{Phys.
  Rev. B}} \href{http://dx.doi.org/10.1103/PhysRevB.89.165118}{{\bf
  \bibinfo{volume}{89}}, \bibinfo{pages}{165118}}
  (\href{http://dx.doi.org/10.1103/PhysRevB.89.165118}{\bibinfo{year}{2014}}).

\bibitem{bonca2012}
\bibinfo{author}{J.~Bon\v{c}a}, \bibinfo{author}{M.~Mierzejewski}, and
  \bibinfo{author}{L.~Vidmar}, \bibinfo{title}{Nonequilibrium propagation and
  decay of a bound pair in driven t-{J} models},
  \bibinfo{journal}{\href{http://dx.doi.org/10.1103/PhysRevLett.109.156404}{Phys.
  Rev. Lett.}} \href{http://dx.doi.org/10.1103/PhysRevLett.109.156404}{{\bf
  \bibinfo{volume}{109}}, \bibinfo{pages}{156404}}
  (\href{http://dx.doi.org/10.1103/PhysRevLett.109.156404}{\bibinfo{year}{2012}}).

\bibitem{iyoda14}
\bibinfo{author}{E.~Iyoda} and \bibinfo{author}{S.~Ishihara},
  \bibinfo{title}{{Transient carrier dynamics in a Mott insulator with
  antiferromagnetic order}},
  \bibinfo{journal}{\href{http://dx.doi.org/10.1103/PhysRevB.89.125126}{Phys.
  Rev. B}} \href{http://dx.doi.org/10.1103/PhysRevB.89.125126}{{\bf
  \bibinfo{volume}{89}}, \bibinfo{pages}{125126}}
  (\href{http://dx.doi.org/10.1103/PhysRevB.89.125126}{\bibinfo{year}{2014}}).

\bibitem{eckstein14b}
\bibinfo{author}{M.~Eckstein} and \bibinfo{author}{P.~Werner},
  \bibinfo{title}{{Ultra-fast photo-carrier relaxation in Mott insulators with
  short-range spin correlations}},
  \bibinfo{journal}{\href{http://dx.doi.org/10.1038/srep21235}{Scientific
  Reports}} \href{http://dx.doi.org/10.1038/srep21235}{{\bf
  \bibinfo{volume}{6}}, \bibinfo{pages}{21235}}
  (\href{http://dx.doi.org/10.1038/srep21235}{\bibinfo{year}{2016}}).

\bibitem{lev2011}
\bibinfo{author}{L.~Vidmar}, \bibinfo{author}{J.~Bon\ifmmode~\check{c}\else
  \v{c}\fi{}a}, \bibinfo{author}{T.~Tohyama}, and \bibinfo{author}{S.~Maekawa},
  \bibinfo{title}{Quantum dynamics of a driven correlated system coupled to
  phonons},
  \bibinfo{journal}{\href{http://dx.doi.org/10.1103/PhysRevLett.107.246404}{Phys.
  Rev. Lett.}} \href{http://dx.doi.org/10.1103/PhysRevLett.107.246404}{{\bf
  \bibinfo{volume}{107}}, \bibinfo{pages}{246404}}
  (\href{http://dx.doi.org/10.1103/PhysRevLett.107.246404}{\bibinfo{year}{2011}}).

\bibitem{kogoj2014}
\bibinfo{author}{J.~Kogoj}, \bibinfo{author}{Z.~Lenar\ifmmode \check{c}\else
  \v{c}\fi{}i\ifmmode~\check{c}\else \v{c}\fi{}},
  \bibinfo{author}{D.~Gole\ifmmode~\check{z}\else \v{z}\fi{}},
  \bibinfo{author}{M.~Mierzejewski},
  \bibinfo{author}{P.~Prelov\ifmmode~\check{s}\else \v{s}\fi{}ek}, and
  \bibinfo{author}{J.~Bon\ifmmode~\check{c}\else \v{c}\fi{}a},
  \bibinfo{title}{Multistage dynamics of the spin-lattice polaron formation},
  \bibinfo{journal}{\href{http://dx.doi.org/10.1103/PhysRevB.90.125104}{Phys.
  Rev. B}} \href{http://dx.doi.org/10.1103/PhysRevB.90.125104}{{\bf
  \bibinfo{volume}{90}}, \bibinfo{pages}{125104}}
  (\href{http://dx.doi.org/10.1103/PhysRevB.90.125104}{\bibinfo{year}{2014}}).

\bibitem{eckstein2013}
\bibinfo{author}{M.~Eckstein} and \bibinfo{author}{P.~Werner},
  \bibinfo{title}{{Photoinduced States in a Mott Insulator}},
  \bibinfo{journal}{\href{http://dx.doi.org/10.1103/PhysRevLett.110.126401}{Phys.
  Rev. Lett.}} \href{http://dx.doi.org/10.1103/PhysRevLett.110.126401}{{\bf
  \bibinfo{volume}{110}}, \bibinfo{pages}{126401}}
  (\href{http://dx.doi.org/10.1103/PhysRevLett.110.126401}{\bibinfo{year}{2013}}).

\bibitem{moritz13}
\bibinfo{author}{B.~Moritz}, \bibinfo{author}{A.~F. Kemper},
  \bibinfo{author}{M.~Sentef}, \bibinfo{author}{T.~P. Devereaux}, and
  \bibinfo{author}{J.~K. Freericks}, \bibinfo{title}{Electron-mediated
  relaxation following ultrafast pumping of strongly correlated materials:
  Model evidence of a correlation-tuned crossover between thermal and
  nonthermal states},
  \bibinfo{journal}{\href{http://dx.doi.org/10.1103/PhysRevLett.111.077401}{Phys.
  Rev. Lett.}} \href{http://dx.doi.org/10.1103/PhysRevLett.111.077401}{{\bf
  \bibinfo{volume}{111}}, \bibinfo{pages}{077401}}
  (\href{http://dx.doi.org/10.1103/PhysRevLett.111.077401}{\bibinfo{year}{2013}}).

\bibitem{rincon14}
\bibinfo{author}{J.~Rinc\'on}, \bibinfo{author}{K.~A. Al-Hassanieh},
  \bibinfo{author}{A.~E. Feiguin}, and \bibinfo{author}{E.~Dagotto},
  \bibinfo{title}{Photoexcitation of electronic instabilities in
  one-dimensional charge-transfer systems},
  \bibinfo{journal}{\href{http://dx.doi.org/10.1103/PhysRevB.90.155112}{Phys.
  Rev. B}} \href{http://dx.doi.org/10.1103/PhysRevB.90.155112}{{\bf
  \bibinfo{volume}{90}}, \bibinfo{pages}{155112}}
  (\href{http://dx.doi.org/10.1103/PhysRevB.90.155112}{\bibinfo{year}{2014}}).

\bibitem{werner14}
\bibinfo{author}{P.~Werner}, \bibinfo{author}{K.~Held}, and
  \bibinfo{author}{M.~Eckstein}, \bibinfo{title}{{Role of impact ionization in
  the thermalization of photoexcited Mott insulators}},
  \bibinfo{journal}{\href{http://dx.doi.org/10.1103/PhysRevB.90.235102}{Phys.
  Rev. B}} \href{http://dx.doi.org/10.1103/PhysRevB.90.235102}{{\bf
  \bibinfo{volume}{90}}, \bibinfo{pages}{235102}}
  (\href{http://dx.doi.org/10.1103/PhysRevB.90.235102}{\bibinfo{year}{2014}}).

\bibitem{mierzejewski2011}
\bibinfo{author}{M.~Mierzejewski}, \bibinfo{author}{L.~Vidmar},
  \bibinfo{author}{J.~Bon\ifmmode~\check{c}\else \v{c}\fi{}a}, and
  \bibinfo{author}{P.~Prelov\ifmmode~\check{s}\else \v{s}\fi{}ek},
  \bibinfo{title}{{Nonequilibrium quantum dynamics of a charge carrier doped
  into a Mott Insulator}},
  \bibinfo{journal}{\href{http://dx.doi.org/10.1103/PhysRevLett.106.196401}{Phys.
  Rev. Lett.}} \href{http://dx.doi.org/10.1103/PhysRevLett.106.196401}{{\bf
  \bibinfo{volume}{106}}, \bibinfo{pages}{196401}}
  (\href{http://dx.doi.org/10.1103/PhysRevLett.106.196401}{\bibinfo{year}{2011}}).

\bibitem{lev2011_1}
\bibinfo{author}{L.~Vidmar}, \bibinfo{author}{J.~Bon\ifmmode~\check{c}\else
  \v{c}\fi{}a}, \bibinfo{author}{M.~Mierzejewski},
  \bibinfo{author}{P.~Prelov\ifmmode~\check{s}\else \v{s}\fi{}ek}, and
  \bibinfo{author}{S.~A. Trugman}, \bibinfo{title}{{Nonequilibrium dynamics of
  the Holstein polaron driven by an external electric field}},
  \bibinfo{journal}{\href{http://dx.doi.org/10.1103/PhysRevB.83.134301}{Phys.
  Rev. B}} \href{http://dx.doi.org/10.1103/PhysRevB.83.134301}{{\bf
  \bibinfo{volume}{83}}, \bibinfo{pages}{134301}}
  (\href{http://dx.doi.org/10.1103/PhysRevB.83.134301}{\bibinfo{year}{2011}}).

\bibitem{golez2012}
\bibinfo{author}{D.~Gole\ifmmode~\check{z}\else \v{z}\fi{}},
  \bibinfo{author}{J.~Bon\ifmmode~\check{c}\else \v{c}\fi{}a}, and
  \bibinfo{author}{L.~Vidmar}, \bibinfo{title}{Dissociation of a
  {H}ubbard-{H}olstein bipolaron driven away from equilibrium by a constant
  electric field},
  \bibinfo{journal}{\href{http://dx.doi.org/10.1103/PhysRevB.85.144304}{Phys.
  Rev. B}} \href{http://dx.doi.org/10.1103/PhysRevB.85.144304}{{\bf
  \bibinfo{volume}{85}}, \bibinfo{pages}{144304}}
  (\href{http://dx.doi.org/10.1103/PhysRevB.85.144304}{\bibinfo{year}{2012}}).

\bibitem{park1986}
\bibinfo{author}{T.~J. Park} and \bibinfo{author}{J.~C. Light},
  \bibinfo{title}{{Unitary quantum time evolution by iterative Lanczos
  reduction}}, \bibinfo{journal}{\href{http://dx.doi.org/10.1063/1.451548}{The
  Journal of Chemical Physics}} \href{http://dx.doi.org/10.1063/1.451548}{{\bf
  \bibinfo{volume}{85}}, \bibinfo{pages}{5870}}
  (\href{http://dx.doi.org/10.1063/1.451548}{\bibinfo{year}{1986}}).

\bibitem{bonca99}
\bibinfo{author}{J.~Bon\v{c}a}, \bibinfo{author}{S.~A. Trugman}, and
  \bibinfo{author}{I.~Batisti\'{c}}, \bibinfo{title}{{Holstein polaron}},
  \bibinfo{journal}{\href{http://dx.doi.org/10.1103/PhysRevB.60.1633}{Phys.
  Rev. B}} \href{http://dx.doi.org/10.1103/PhysRevB.60.1633}{{\bf
  \bibinfo{volume}{60}}, \bibinfo{pages}{1633}}
  (\href{http://dx.doi.org/10.1103/PhysRevB.60.1633}{\bibinfo{year}{1999}}).

\bibitem{ku2002}
\bibinfo{author}{L.-C. Ku}, \bibinfo{author}{S.~A. Trugman}, and
  \bibinfo{author}{J.~Bon\v{c}a}, \bibinfo{title}{{Dimensionality effects on
  the Holstein polaron}},
  \bibinfo{journal}{\href{http://dx.doi.org/10.1103/PhysRevB.65.174306}{Phys.
  Rev. B}} \href{http://dx.doi.org/10.1103/PhysRevB.65.174306}{{\bf
  \bibinfo{volume}{65}}, \bibinfo{pages}{174306}}
  (\href{http://dx.doi.org/10.1103/PhysRevB.65.174306}{\bibinfo{year}{2002}}).

\bibitem{jaklic00}
\bibinfo{author}{J.~Jakli{\v{c}}} and \bibinfo{author}{P.~Prelov{\v{s}}ek},
  \bibinfo{title}{Finite-temperature properties of doped antiferromagnets},
  \bibinfo{journal}{\href{http://dx.doi.org/10.1080/000187300243381}{Adv.
  Phys.}} \href{http://dx.doi.org/10.1080/000187300243381}{{\bf
  \bibinfo{volume}{49}}, \bibinfo{pages}{1}}
  (\href{http://dx.doi.org/10.1080/000187300243381}{\bibinfo{year}{2000}}).

\bibitem{our2013}
\bibinfo{author}{M.~Mierzejewski}, \bibinfo{author}{T.~Prosen},
  \bibinfo{author}{D.~Crivelli}, and
  \bibinfo{author}{P.~Prelov\ifmmode~\check{s}\else \v{s}\fi{}ek},
  \bibinfo{title}{Eigenvalue statistics of reduced density matrix during
  driving and relaxation},
  \bibinfo{journal}{\href{http://dx.doi.org/10.1103/PhysRevLett.110.200602}{Phys.
  Rev. Lett.}} \href{http://dx.doi.org/10.1103/PhysRevLett.110.200602}{{\bf
  \bibinfo{volume}{110}}, \bibinfo{pages}{200602}}
  (\href{http://dx.doi.org/10.1103/PhysRevLett.110.200602}{\bibinfo{year}{2013}}).

\bibitem{sedlmayr13}
\bibinfo{author}{N.~Sedlmayr}, \bibinfo{author}{J.~Ren},
  \bibinfo{author}{F.~Gebhard}, and \bibinfo{author}{J.~Sirker},
  \bibinfo{title}{Closed and open system dynamics in a fermionic chain with a
  microscopically specified bath: Relaxation and thermalization},
  \bibinfo{journal}{\href{http://dx.doi.org/10.1103/PhysRevLett.110.100406}{Phys.
  Rev. Lett.}} \href{http://dx.doi.org/10.1103/PhysRevLett.110.100406}{{\bf
  \bibinfo{volume}{110}}, \bibinfo{pages}{100406}}
  (\href{http://dx.doi.org/10.1103/PhysRevLett.110.100406}{\bibinfo{year}{2013}}).

\bibitem{fagotti13}
\bibinfo{author}{M.~Fagotti} and \bibinfo{author}{F.~H.~L. Essler},
  \bibinfo{title}{Reduced density matrix after a quantum quench},
  \bibinfo{journal}{\href{http://dx.doi.org/10.1103/PhysRevB.87.245107}{Phys.
  Rev. B}} \href{http://dx.doi.org/10.1103/PhysRevB.87.245107}{{\bf
  \bibinfo{volume}{87}}, \bibinfo{pages}{245107}}
  (\href{http://dx.doi.org/10.1103/PhysRevB.87.245107}{\bibinfo{year}{2013}}).

\bibitem{mahan2000}
\bibinfo{author}{G.~D. Mahan}, {\em \bibinfo{title}{Many-Particle Physics}\/}
  (\bibinfo{publisher}{Springer, New York}, \bibinfo{year}{2000}).

\bibitem{eckstein08}
\bibinfo{author}{M.~Eckstein} and \bibinfo{author}{M.~Kollar},
  \bibinfo{title}{Theory of time-resolved optical spectroscopy on correlated
  electron systems},
  \bibinfo{journal}{\href{http://dx.doi.org/10.1103/PhysRevB.78.205119}{Phys.
  Rev. B}} \href{http://dx.doi.org/10.1103/PhysRevB.78.205119}{{\bf
  \bibinfo{volume}{78}}, \bibinfo{pages}{205119}}
  (\href{http://dx.doi.org/10.1103/PhysRevB.78.205119}{\bibinfo{year}{2008}}).

\bibitem{unterhinninghofen08}
\bibinfo{author}{J.~Unterhinninghofen}, \bibinfo{author}{D.~Manske}, and
  \bibinfo{author}{A.~Knorr}, \bibinfo{title}{Theory of ultrafast
  nonequilibrium dynamics in $d$-wave superconductors},
  \bibinfo{journal}{\href{http://dx.doi.org/10.1103/PhysRevB.77.180509}{Phys.
  Rev. B}} \href{http://dx.doi.org/10.1103/PhysRevB.77.180509}{{\bf
  \bibinfo{volume}{77}}, \bibinfo{pages}{180509}}
  (\href{http://dx.doi.org/10.1103/PhysRevB.77.180509}{\bibinfo{year}{2008}}).

\bibitem{eckstein10}
\bibinfo{author}{M.~Eckstein}, \bibinfo{author}{M.~Kollar}, and
  \bibinfo{author}{P.~Werner}, \bibinfo{title}{{Interaction quench in the
  Hubbard model: Relaxation of the spectral function and the optical
  conductivity}},
  \bibinfo{journal}{\href{http://dx.doi.org/10.1103/PhysRevB.81.115131}{Phys.
  Rev. B}} \href{http://dx.doi.org/10.1103/PhysRevB.81.115131}{{\bf
  \bibinfo{volume}{81}}, \bibinfo{pages}{115131}}
  (\href{http://dx.doi.org/10.1103/PhysRevB.81.115131}{\bibinfo{year}{2010}}).

\bibitem{kanamori10}
\bibinfo{author}{Y.~Kanamori}, \bibinfo{author}{H.~Matsueda}, and
  \bibinfo{author}{S.~Ishihara}, \bibinfo{title}{Numerical study of
  photoinduced dynamics in a double-exchange model},
  \bibinfo{journal}{\href{http://dx.doi.org/10.1103/PhysRevB.82.115101}{Phys.
  Rev. B}} \href{http://dx.doi.org/10.1103/PhysRevB.82.115101}{{\bf
  \bibinfo{volume}{82}}, \bibinfo{pages}{115101}}
  (\href{http://dx.doi.org/10.1103/PhysRevB.82.115101}{\bibinfo{year}{2010}}).

\bibitem{wall11}
\bibinfo{author}{S.~Wall}, \bibinfo{author}{D.~Brida}, \bibinfo{author}{S.~R.
  Clark}, \bibinfo{author}{H.~P. Ehrke}, \bibinfo{author}{D.~Jaksch},
  \bibinfo{author}{A.~Ardavan}, \bibinfo{author}{S.~Bonora},
  \bibinfo{author}{H.~Uemura}, \bibinfo{author}{Y.~Takahashi},
  \bibinfo{author}{T.~Hasegawa}, \bibinfo{author}{H.~Okamoto},
  \bibinfo{author}{G.~Cerullo}, and \bibinfo{author}{A.~Cavalleri},
  \bibinfo{title}{{Quantum interference between charge excitation paths in a
  solid-state Mott insulator}},
  \bibinfo{journal}{\href{http://dx.doi.org/10.1038/nphys1831}{Nature Physics}}
  \href{http://dx.doi.org/10.1038/nphys1831}{{\bf \bibinfo{volume}{7}},
  \bibinfo{pages}{114}}
  (\href{http://dx.doi.org/10.1038/nphys1831}{\bibinfo{year}{2011}}).

\bibitem{shimizu11}
\bibinfo{author}{A.~Shimizu} and \bibinfo{author}{T.~Yuge}, \bibinfo{title}{Sum
  rules and asymptotic behaviors for optical conductivity of nonequilibrium
  many-electron systems},
  \bibinfo{journal}{\href{http://dx.doi.org/10.1143/JPSJ.80.093706}{J. Phys.
  Soc. Jpn.}} \href{http://dx.doi.org/10.1143/JPSJ.80.093706}{{\bf
  \bibinfo{volume}{80}}, \bibinfo{pages}{093706}}
  (\href{http://dx.doi.org/10.1143/JPSJ.80.093706}{\bibinfo{year}{2011}}).

\bibitem{zala2014}
\bibinfo{author}{Z.~Lenar\ifmmode \check{c}\else
  \v{c}\fi{}i\ifmmode~\check{c}\else \v{c}\fi{}},
  \bibinfo{author}{D.~Gole\ifmmode~\check{z}\else \v{z}\fi{}},
  \bibinfo{author}{J.~Bon\ifmmode~\check{c}\else \v{c}\fi{}a}, and
  \bibinfo{author}{P.~Prelov\ifmmode~\check{s}\else \v{s}\fi{}ek},
  \bibinfo{title}{Optical response of highly excited particles in a strongly
  correlated system},
  \bibinfo{journal}{\href{http://dx.doi.org/10.1103/PhysRevB.89.125123}{Phys.
  Rev. B}} \href{http://dx.doi.org/10.1103/PhysRevB.89.125123}{{\bf
  \bibinfo{volume}{89}}, \bibinfo{pages}{125123}}
  (\href{http://dx.doi.org/10.1103/PhysRevB.89.125123}{\bibinfo{year}{2014}}).

\bibitem{lu15}
\bibinfo{author}{H.~Lu}, \bibinfo{author}{C.~Shao},
  \bibinfo{author}{J.~Bon\ifmmode~\check{c}\else \v{c}\fi{}a},
  \bibinfo{author}{D.~Manske}, and \bibinfo{author}{T.~Tohyama},
  \bibinfo{title}{Photoinduced in-gap excitations in the one-dimensional
  extended hubbard model},
  \bibinfo{journal}{\href{http://dx.doi.org/10.1103/PhysRevB.91.245117}{Phys.
  Rev. B}} \href{http://dx.doi.org/10.1103/PhysRevB.91.245117}{{\bf
  \bibinfo{volume}{91}}, \bibinfo{pages}{245117}}
  (\href{http://dx.doi.org/10.1103/PhysRevB.91.245117}{\bibinfo{year}{2015}}).

\bibitem{shao15}
\bibinfo{author}{C.~Shao}, \bibinfo{author}{T.~Tohyama}, \bibinfo{author}{H.-G.
  Luo}, and \bibinfo{author}{H.~Lu}, \bibinfo{title}{Numerical method to
  compute optical conductivity based on pump-probe simulations},
  \bibinfo{journal}{\href{http://dx.doi.org/10.1103/PhysRevB.93.195144}{Phys.
  Rev. B}} \href{http://dx.doi.org/10.1103/PhysRevB.93.195144}{{\bf
  \bibinfo{volume}{93}}, \bibinfo{pages}{195144}}
  (\href{http://dx.doi.org/10.1103/PhysRevB.93.195144}{\bibinfo{year}{2016}}).

\bibitem{wright14}
\bibinfo{author}{T.~M. Wright}, \bibinfo{author}{M.~Rigol},
  \bibinfo{author}{M.~J. Davis}, and \bibinfo{author}{K.~V. Kheruntsyan},
  \bibinfo{title}{Nonequilibrium dynamics of one-dimensional hard-core anyons
  following a quench: Complete relaxation of one-body observables},
  \bibinfo{journal}{\href{http://dx.doi.org/10.1103/PhysRevLett.113.050601}{Phys.
  Rev. Lett.}} \href{http://dx.doi.org/10.1103/PhysRevLett.113.050601}{{\bf
  \bibinfo{volume}{113}}, \bibinfo{pages}{050601}}
  (\href{http://dx.doi.org/10.1103/PhysRevLett.113.050601}{\bibinfo{year}{2014}}).

\bibitem{vidmar16}
\bibinfo{author}{L.~Vidmar} and \bibinfo{author}{M.~Rigol},
  \bibinfo{title}{Generalized gibbs ensemble in integrable lattice models},
  \bibinfo{journal}{\href{http://dx.doi.org/10.1088/1742-5468/2016/06/064007}{J.
  Stat. Mech.}} \href{http://dx.doi.org/10.1088/1742-5468/2016/06/064007}{{\bf
  \bibinfo{volume}{{\rm (2016)}}}, \bibinfo{pages}{064007}}.

\bibitem{perfetti07}
\bibinfo{author}{L.~Perfetti}, \bibinfo{author}{P.~A. Loukakos},
  \bibinfo{author}{M.~Lisowski}, \bibinfo{author}{U.~Bovensiepen},
  \bibinfo{author}{H.~Eisaki}, and \bibinfo{author}{M.~Wolf},
  \bibinfo{title}{Ultrafast electron relaxation in superconducting
  $\mathrm{Bi}_{2}\mathrm{Sr}_{2}\mathrm{CaCu}_{2}\mathrm{O}_{8+\ensuremath{\delta}}$
  by time-resolved photoelectron spectroscopy},
  \bibinfo{journal}{\href{http://dx.doi.org/10.1103/PhysRevLett.99.197001}{Phys.
  Rev. Lett.}} \href{http://dx.doi.org/10.1103/PhysRevLett.99.197001}{{\bf
  \bibinfo{volume}{99}}, \bibinfo{pages}{197001}}
  (\href{http://dx.doi.org/10.1103/PhysRevLett.99.197001}{\bibinfo{year}{2007}}).

\bibitem{bovensiepen12}
\bibinfo{author}{U.~Bovensiepen} and \bibinfo{author}{P.~Kirchmann},
  \bibinfo{title}{Elementary relaxation processes investigated by femtosecond
  photoelectron spectroscopy of two-dimensional materials},
  \bibinfo{journal}{\href{http://dx.doi.org/10.1002/lpor.201000035}{Laser
  Photonics Rev.}} \href{http://dx.doi.org/10.1002/lpor.201000035}{{\bf
  \bibinfo{volume}{6}}, \bibinfo{pages}{589}}
  (\href{http://dx.doi.org/10.1002/lpor.201000035}{\bibinfo{year}{2012}}).

\bibitem{smallwood12b}
\bibinfo{author}{C.~L. Smallwood}, \bibinfo{author}{C.~Jozwiak},
  \bibinfo{author}{W.~Zhang}, and \bibinfo{author}{A.~Lanzara},
  \bibinfo{title}{An ultrafast angle-resolved photoemission apparatus for
  measuring complex materials},
  \bibinfo{journal}{\href{http://dx.doi.org/http://dx.doi.org/10.1063/1.4772070}{Review
  of Scientific Instruments}}
  \href{http://dx.doi.org/http://dx.doi.org/10.1063/1.4772070}{{\bf
  \bibinfo{volume}{83}}, \bibinfo{eid}{123904}}
  (\href{http://dx.doi.org/http://dx.doi.org/10.1063/1.4772070}{\bibinfo{year}{2012}}).

\bibitem{piovera15}
\bibinfo{author}{C.~Piovera}, \bibinfo{author}{Z.~Zhang},
  \bibinfo{author}{M.~d'Astuto}, \bibinfo{author}{A.~Taleb-Ibrahimi},
  \bibinfo{author}{E.~Papalazarou}, \bibinfo{author}{M.~Marsi},
  \bibinfo{author}{Z.~Z. Li}, \bibinfo{author}{H.~Raffy}, and
  \bibinfo{author}{L.~Perfetti}, \bibinfo{title}{Quasiparticle dynamics in
  high-temperature superconductors far from equilibrium: An indication of
  pairing amplitude without phase coherence},
  \bibinfo{journal}{\href{http://dx.doi.org/10.1103/PhysRevB.91.224509}{Phys.
  Rev. B}} \href{http://dx.doi.org/10.1103/PhysRevB.91.224509}{{\bf
  \bibinfo{volume}{91}}, \bibinfo{pages}{224509}}
  (\href{http://dx.doi.org/10.1103/PhysRevB.91.224509}{\bibinfo{year}{2015}}).

\bibitem{rameau15}
\bibinfo{author}{J.~D. Rameau}, \bibinfo{author}{S.~Freutel},
  \bibinfo{author}{M.~A. Sentef}, \bibinfo{author}{A.~F. Kemper},
  \bibinfo{author}{J.~K. Freericks}, \bibinfo{author}{I.~Avigo},
  \bibinfo{author}{M.~Ligges}, \bibinfo{author}{L.~Rettig},
  \bibinfo{author}{Y.~Yoshida}, \bibinfo{author}{H.~Eisaki},
  \bibinfo{author}{J.~Schneeloch}, \bibinfo{author}{R.~D. Zhong},
  \bibinfo{author}{Z.~J. Xu}, \bibinfo{author}{G.~D. Gu},
  \bibinfo{author}{P.~D. Johnson}, and \bibinfo{author}{U.~Bovensiepen},
  \bibinfo{title}{Time-resolved boson emission in the excitation spectrum of
  $\mathrm{Bi}_{2}\mathrm{Sr}_{2}\mathrm{CaCu}_{2}\mathrm{O}_{8+\ensuremath{\delta}}$},
  \href{http://arxiv.org/abs/1505.07055}{\bibinfo{howpublished}{arXiv:1505.07055}}.

\bibitem{prelovvsek2013strongly}
\bibinfo{author}{P.~Prelov{\v{s}}ek} and \bibinfo{author}{J.~Bon{\v{c}}a}, {\em
  \bibinfo{title}{Ground state and finite temperature Lanczos methods}\/},
  volume \bibinfo{volume}{176} of \bibinfo{series}{Springer Series in
  Solid-State Sciences}, chapter~\bibinfo{chapter}{1}, \bibinfo{pages}{1--30}
  (\bibinfo{publisher}{Springer-Verlag Berlin Heidelberg},
  \bibinfo{year}{2013}).

\bibitem{vidmar10}
\bibinfo{author}{L.~Vidmar}, \bibinfo{author}{J.~Bon\v{c}a}, and
  \bibinfo{author}{S.~A. Trugman}, \bibinfo{title}{{Emergence of states in the
  phonon spectral function of the Holstein polaron below and above the
  one-phonon continuum}},
  \bibinfo{journal}{\href{http://dx.doi.org/10.1103/PhysRevB.82.104304}{Phys.
  Rev. B}} \href{http://dx.doi.org/10.1103/PhysRevB.82.104304}{{\bf
  \bibinfo{volume}{82}}, \bibinfo{pages}{104304}}
  (\href{http://dx.doi.org/10.1103/PhysRevB.82.104304}{\bibinfo{year}{2010}}).

\bibitem{brockt15}
\bibinfo{author}{C.~Brockt}, \bibinfo{author}{F.~Dorfner},
  \bibinfo{author}{L.~Vidmar}, \bibinfo{author}{F.~Heidrich-Meisner}, and
  \bibinfo{author}{E.~Jeckelmann}, \bibinfo{title}{Matrix-product-state method
  with a dynamical local basis optimization for bosonic systems out of
  equilibrium},
  \bibinfo{journal}{\href{http://dx.doi.org/10.1103/PhysRevB.92.241106}{Phys.
  Rev. B}} \href{http://dx.doi.org/10.1103/PhysRevB.92.241106}{{\bf
  \bibinfo{volume}{92}}, \bibinfo{pages}{241106}}
  (\href{http://dx.doi.org/10.1103/PhysRevB.92.241106}{\bibinfo{year}{2015}}).

\end{thebibliography}

\end{document}